 \documentclass[final,3p,times,sort&compress]{elsarticle}
\usepackage{multirow,setspace,times,amssymb,amsmath,graphicx,color,rotating,subfigure,url}
\usepackage{lineno}
\usepackage{natbib}

\journal{Physica A} %% change this journal name and put the correct one

\begin{document}

\begin{frontmatter}

\title{Scaling and memory in the non-poisson process of limit order cancelation}
\author[BS,RCE,SS]{Xiao-Hui Ni}
\author[BS,RCE,SS]{Zhi-Qiang Jiang}
\author[BS,RCE,SS]{Gao-Feng Gu}
\author[BS,RCE,RCSE]{Fei Ren}
\author[SZSC]{Wei Chen}
\author[BS,RCE,SS,RCSE,RCFE]{Wei-Xing Zhou \corref{cor}}
\cortext[cor]{Corresponding author. Address: 130 Meilong Road, P.O. Box 114, School of Business, East China University of Science and Technology, Shanghai 200237, China, Phone: +86 21 64253634, Fax: +86 21 64253152.}
\ead{wxzhou@ecust.edu.cn} %

\address[BS]{School of Business, East China University of Science and Technology, Shanghai 200237, China}
\address[RCE]{Research Center for Econophysics, East China University of Science and Technology, Shanghai 200237, China}
\address[SS]{School of Science, East China University of Science and Technology, Shanghai 200237, China}
\address[RCSE]{Engineering Research Center of Process Systems Engineering (Ministry of Education), East China University of Science and Technology, Shanghai 200237, China}
\address[SZSC]{Shenzhen Stock Exchange, 5045 Shennan East Road, Shenzhen 518010, China}
\address[RCFE]{Research Center on Fictitious Economics \& Data Science, Chinese Academy of Sciences, Beijing 100080, China}

\begin{abstract}
The order submission and cancelation processes are two crucial aspects in the price formation of stocks traded in order-driven markets. We investigate the dynamics of order cancelation by studying the statistical properties of inter-cancelation durations defined as the waiting times between consecutive order cancelations of 22 liquid stocks traded on the Shenzhen Stock Exchange of China in year 2003. Three types of cancelations are considered including cancelation of any limit orders, of buy limit orders and of sell limit orders. We find that the distributions of the inter-cancelation durations of individual stocks can be well modeled by Weibulls  for each type of cancelation and the distributions of rescaled durations of each type of cancelations exhibit a scaling behavior for different stocks. Complex intraday patterns are also unveiled in the inter-cancelation durations. The detrended fluctuation analysis (DFA) and the multifractal DFA show that the inter-cancelation durations possess long-term memory and multifractal nature, which are not influenced by the intraday patterns. No clear crossover phenomenon is observed in the detrended fluctuation functions with respect to the time scale. These findings indicate that the cancelation of limit orders is a non-Poisson process, which has potential worth in the construction of order-driven market models.
\end{abstract}


\end{frontmatter}

\section{Introduction}
\label{S1:intro}

Order submission and cancelation are two central processes in the price formation of stocks traded in order-driven markets. Understanding their statistical regularities are crucial in the study of stock market microstructure theory and the construction of order-driven models \cite{Mike-Farmer-2008-JEDC,Gu-Zhou-2009-EPJB,Gu-Zhou-2009-EPL}. For limit orders, there are three attributes: order direction (or order sign indicating buy/sell), order price, and order size. The statistical properties of these quantities in the order submission process have been extensively studied including the long memory of order signs \cite{Lillo-Farmer-2004-SNDE,Mike-Farmer-2008-JEDC}, the distribution of relative prices \cite{Zovko-Farmer-2002-QF,Bouchaud-Mezard-Potters-2002-QF,Potters-Bouchaud-2003-PA,Maskawa-2007-PA,Mike-Farmer-2008-JEDC,Ranaldo-2004-JFinM,Lillo-2007-EPJB,Gu-Chen-Zhou-2008b-PA,Gu-Ren-Ni-Chen-Zhou-2010-PA}, the long memory of relative prices known as the``diagonal effect''
\cite{Biais-Hillion-Spatt-1995-JF,Zovko-Farmer-2002-QF,Gu-Zhou-2009-EPL}, and the power-law distribution and long memory of order sizes and trading volumes \cite{Gopikrishnan-Plerou-Gabaix-Stanley-2000-PRE,Plerou-Gopikrishnan-Gabaix-Amaral-Stanley-2001-QF,Maslov-Mills-2001-PA,Tsallis-Anteneodo-Borland-Osorio-2003-PA,Osorio-Borland-Tsallis-2004,Plerou-Gopikrishnan-Gabaix-Stanley-2004-QF,Farmer-Lillo-2004-QF,Queiros-2005-EPL,deSouza-Moyano-Queiros-2006-EPJB,Queiros-Moyano-deSouza-Tsallis-2007-EPJB,Plerou-Stanley-2007-PRE,Eisler-Kertesz-2006-EPJB,Eisler-Kertesz-2007-PA,Lee-Lee-2007-PA,Zhou-2007-XXX,Racz-Eisler-Kertesz-2009-PRE,Plerou-Stanley-2009-PRE,Qiu-Zhong-Chen-Wu-2009-PA,Mu-Chen-Kertesz-Zhou-2009-EPJB}.
In contrast, the statistical regularities of the order cancelation process are less studied. When constructing the empirical behavioral model of order-driven stock markets, the conditional probability of order cancelation on three factors has been determined \cite{Mike-Farmer-2008-JEDC}. In this work, we attempt to understand the order cancelation dynamics by investigating the inter-cancelation durations that are the waiting time between consecutive cancelations.

There are numerous studies on the waiting time distributions of diverse financial quantities. Generally speaking, the waiting time is defined as the time interval between two successive financial events. When the financial events are defined, one is able to determine the series of waiting times. One importance topic is about the return intervals (or recurrence intervals). On the one hand, the recurrence intervals between financial volatilities
exceeding a certain threshold $q$ have been carefully studied, and numerous phenomena have been unveiled
\cite{Kaizoji-Kaizoji-2004a-PA,Yamasaki-Muchnik-Havlin-Bunde-Stanley-2005-PNAS,Wang-Yamasaki-Havlin-Stanley-2006-PRE,Lee-Lee-Rikvold-2006-JKPS,Wang-Weber-Yamasaki-Havlin-Stanley-2007-EPJB,VodenskaChitkushev-Wang-Weber-Yamasaki-Havlin-Stanley-2008-EPJB,Jung-Wang-Havlin-Kaizoji-Moon-Stanley-2008-EPJB,Wang-Yamasaki-Havlin-Stanley-2008-PRE,Qiu-Guo-Chen-2008-PA,Ren-Zhou-2008-EPL,Ren-Guo-Zhou-2009-PA,Ren-Gu-Zhou-2009-PA}.
On the other hand, econophysicists have devoted to the study of recurrence intervals between large positive or negative price returns, which has important implications on risk estimation
\cite{Yamasaki-Muchnik-Havlin-Bunde-Stanley-2006-inPFE,Bogachev-Eichner-Bunde-2007-PRL,Bogachev-Bunde-2008-PRE,Bogachev-Bunde-2009-PRE,Ren-Zhou-2010-NJP}.

Another important topic is about the intertrade durations, which are defined as the waiting times between consecutive transactions of an equity. The importance of this topic is due to the fact that intertrade durations contain information contents of trading activity and have crucial relevance to the microstructure theory \cite{Diamond-Verrecchia-1987-JFE,Easley-OHara-1992-JF,Dufour-Engle-2000-JF}. In mainstream finance, the celebrated autoregressive conditional duration (ACD) model was proposed to model the intertrade durations with temporal correlation and other financial variables \cite{Engle-Russell-1998-Em,Engle-2000-Em}. Alternatively, in the econophysics community, the continuous-time random walk (CTRW) formalism has been adopted to deal with the intertrade durations and price dynamics
\cite{Scalas-Gorenflo-Mainardi-2000-PA,Mainardi-Raberto-Gorenflo-Scalas-2000-PA,Masoliver-Montero-Weiss-2003-PRE,Scalas-2006-PA,Masoliver-Montero-Perello-Weiss-2006-JEBO}.
Empirical investigations of the intertrade durations from different equity markets report that the probability distributions might be described by
power laws \cite{Sabatelli-Keating-Dudley-Richmond-2002-EPJB,Yoon-Choi-Lee-Yum-Kim-2006-PA}, modified power laws \cite{Masoliver-Montero-Weiss-2003-PRE,Masoliver-Montero-Perello-Weiss-2006-JEBO}, stretched exponentials (or Weibulls)
\cite{Raberto-Scalas-Mainardi-2002-PA,Bartiromo-2004-PRE,Ivanov-Yuen-Podobnik-Lee-2004-PRE,Eisler-Kertesz-2006-EPJB,Sazuka-2007-PA,Poloti-Scalas-2008-PA,Jiang-Chen-Zhou-2008-PA}, stretched exponentials followed by power laws \cite{Kim-Yoon-2003-Fractals,Kim-Yoon-Kim-Lee-Scalas-2007-JKPS}, implying that the transaction process is non-Poisson.
However, statistical tests reject the hypothesis that the intertrade durations are distributed according to an exponential
\cite{Scalas-Gorenflo-Luckock-Mainardi-Mantelli-Raberto-2004-QF,Scalas-Gorenflo-Luckock-Mainardi-Mantelli-Raberto-2005-FL} or a power law \cite{Poloti-Scalas-2008-PA}. Hence, the Weibull seems to be the very form of intertrade duration distribution \cite{Raberto-Scalas-Mainardi-2002-PA,Bartiromo-2004-PRE,Ivanov-Yuen-Podobnik-Lee-2004-PRE,Sazuka-2007-PA,Poloti-Scalas-2008-PA,Jiang-Chen-Zhou-2008-PA}. In addition, a scaling behavior can be observed in the distributions of rescaled intertrade durations \cite{Ivanov-Yuen-Podobnik-Lee-2004-PRE,Poloti-Scalas-2008-PA,Jiang-Chen-Zhou-2008-PA}, which is however less conclusive \cite{Eisler-Kertesz-2006-EPJB}.

The cancelation of limit orders has certain impact on the price formation. Especially, when the orders at the best bid or ask prices on the order book are canceled, the bid-ask spread widens and the mid-price changes. In some order-driven models, the cancelation is assumed to follow a Poisson process \cite{Daniels-Farmer-Gillemot-Iori-Smith-2003-PRL}. Although this is a good approximation, we find in this work that the inter-cancelation durations are distributed according to a Weibull rather than an exponential and long-term correlated. These findings show that the cancelation of limit orders is a non-Poisson process, which is useful in constructing more realistic order-driven market models.

The rest of this paper is organized as follows. In Section \ref{S2:dataset}, we briefly describe the data sets investigated and the basic statistics of limit order cancelations. Section \ref{S3:dist} investigates the empirical distributions of the inter-cancelation duration. In section \ref{S1:Memory}, the temporal correlations and the multifractal nature of the inter-cancelation durations are studied based on the (multifractal) detrended fluctuation analysis. Section \ref{S5:conclusion} summarizes and concludes.

\section{Data description}
\label{S2:dataset}

Shenzhen Stock Exchange (SZSE) is an order-driven market. There were
three different periods on each trading day in the SZSE before July
1, 2006, namely, the opening call action (9:15 am to 9:25 am), the
cooling period (9:25 am to 9:30 am), and the continuous double
auction (9:30 am to 11:30 am and 13:00 pm to 15:00 pm). More
information of interest about this market can be found in the literature
\cite{Gu-Chen-Zhou-2007-EPJB,Gu-Chen-Zhou-2008a-PA,Gu-Chen-Zhou-2008b-PA}.
This study is based on the ultrahigh-frequency data of 22 liquid stocks traded
on the SZSE, which contain all order placements and cancelations in
the whole year 2003. Since part of the data of stock 000002, 000027,
000063, 000088, 000089 are missing, the data from 39-th to 167-th
trading days of these stocks are discarded. We only take into consideration the cancelations occurring in the
continuous double auction, and the interval
across the pausing period is excluded. Since the orders not executed
during the continuous double auction period will be canceled
automatically by the system, no inter-cancelation duration will be
calculated overnight either.

Inter-cancelation duration is defined as the waiting time between
two consecutive cancelations (in units of second). The resolution of time is 0.01 second. When the $(i+1)$-th cancelation order arrives, the value
$\tau_i=t_{i+1}-t_i$ is recorded as the $i$-th inter-cancelation
duration, where $t_i$ is the time when the $i$-th cancelation
occurs. The cancelations in the market can be classified into two types
based on their order directions, namely the cancelations of buy orders (buyer-initialed cancelations) and the cancelations of sell orders (seller-initialed cancelations). By analogy with the definition of $\tau_i$, the duration $\tau^b_i$
($\tau^s_i$) between successive cancelations of buy (sell) orders is defined as
\begin{equation}
 \tau^u_i=t^u_{i+1}-t^u_i~,~u\in\{b,s\}, \label{Eq:tau:buy:sell}
\end{equation}
where $t^b_i$ ($t^s_i$) is the time stamp when the $i$-th
buyer-initialed (seller-initialed) cancelation occurs. There are
less than 2\% cancelations occurring at the same time
for each stock.

Table \ref{Tb:NoC} lists the number of cancelations ($N,N^b,N^s$), the number of simultaneously happened
cancelations ($N_0,N_0^b,N_0^s$) and the average inter-cancelation duration
($\langle\tau\rangle, \langle\tau^b\rangle, \langle\tau^s\rangle$) for the cancelation of all orders, buy orders, and sell orders for the 22 Chinese stocks under investigation. We note that there are several trivial relations in this table stating that (1) the total number of cancelation is the sum of the cancelation numbers of buy orders and sell orders
\begin{subequations}
\begin{equation}
 N = N^b+N^s,
\end{equation}
(2) there are less simultaneous cancelations of distinct buy/sell limit orders than all limit orders
\begin{equation}
 N_0 > N^b_0+N^s_0,
\end{equation}
and (3) the mean duration of consecutive cancelations of buy orders or sell orders is greater than the mean duration of all cancelations
\begin{equation}
 \langle\tau^b\rangle > \langle\tau\rangle ~~{\rm{and}}~~ \langle\tau^s\rangle > \langle\tau\rangle.
\end{equation}
\end{subequations}
These relations hold for all stocks.

\begin{table}[htp]
\centering
\caption{Descriptive statistics of the 22 Chinese stocks
studied over the whole year 2003. Since part of the data of stock
000002, 000027, 000063, 000088, 000089 are missing, the data from
39-th to 167-th trading days of these stocks are not taken into
consideration. $N$ is the number of cancelations, $N_0$ is the
number of simultaneous cancelations, and
$\langle\tau\rangle$ is the average inter-cancelation duration for
the 22 Chinese stocks.}
\medskip
\label{Tb:NoC}
\begin{tabular}{lrrrrrcrrc}
 \hline \hline
  \multicolumn{1}{c}{Code} &
  \multicolumn{3}{c}{\begin{tabular}{rrr} \multicolumn{3}{c}{~~~~All cancelation}~~~~ \\ \hline $N$ & ~~~~~~$N_0$ & $\langle\tau\rangle$
  \end{tabular}} &
  \multicolumn{3}{c}{\begin{tabular}{rrc} \multicolumn{3}{c}{Buyer-initiated cancelation} \\ \hline $N^b$ & ~~~~~~~~$N^b_0$ & $\langle\tau^b\rangle$
  \end{tabular}} &
  \multicolumn{3}{c}{\begin{tabular}{rrc} \multicolumn{3}{c}{Seller-initiated cancelation} \\ \hline $N^s$ & ~~~~~~~~$N^s_0$ & $\langle\tau^s\rangle$
  \end{tabular}} \\
  \hline
    000001 &    591944 &     11178 &    5.77 &    317015 &      5141 &   10.75 &    274929 &      4012 &   12.38 \\
    000002 &     78378 &      1220 &   20.05 &     34577 &       726 &   45.14 &     43801 &       391 &   35.66 \\
    000009 &    371225 &      3440 &    9.27 &    183804 &      1183 &   18.68 &    187421 &      1347 &   18.30 \\
    000012 &    221023 &      1853 &   15.37 &    114662 &       659 &   29.40 &    106361 &       849 &   31.82 \\
    000016 &    119408 &      1069 &   28.58 &     60219 &       444 &   56.34 &     59189 &       525 &   57.29 \\
    000021 &    310027 &      3193 &   11.05 &    157174 &      1633 &   21.75 &    152853 &      1068 &   22.32 \\
    000024 &     88187 &      1068 &   38.11 &     42593 &       741 &   78.17 &     45594 &       268 &   73.01 \\
    000027 &     71705 &      1006 &   21.74 &     33058 &       545 &   46.77 &     38647 &       392 &   40.09 \\
    000063 &     60496 &      1136 &   26.05 &     25681 &       660 &   60.68 &     34815 &       414 &   44.98 \\
    000066 &    216984 &      1554 &   15.77 &    110289 &       643 &   30.90 &    106695 &       577 &   31.82 \\
    000088 &     13778 &       625 &  112.29 &      6861 &       410 &  220.82 &      6917 &       212 &  219.50 \\
    000089 &     43893 &      1129 &   35.30 &     20909 &       486 &   73.45 &     22984 &       605 &   66.98 \\
    000429 &     73173 &       527 &   46.15 &     36999 &       311 &   90.50 &     36174 &       164 &   92.15 \\
    000488 &     66439 &      1424 &   51.02 &     32585 &       583 &  102.72 &     33854 &       789 &   99.24 \\
    000539 &     54037 &      3790 &   61.95 &     26950 &      3000 &  120.76 &     27087 &       748 &  122.47 \\
    000541 &     39562 &       580 &   85.82 &     19715 &       218 &  169.87 &     19847 &       351 &  167.93 \\
    000550 &    252471 &      4696 &   13.57 &    122865 &      3236 &   27.75 &    129606 &      1042 &   26.25 \\
    000581 &     56726 &      1798 &   59.56 &     27236 &      1284 &  122.47 &     29490 &       483 &  112.81 \\
    000625 &    255080 &      5399 &   13.23 &    123361 &      3263 &   27.24 &    131719 &      1753 &   25.49 \\
    000709 &    129721 &      1267 &   26.17 &     65704 &       602 &   51.20 &     64017 &       517 &   52.61 \\
    000720 &     30767 &      3805 &  103.55 &     16558 &      2490 &  175.61 &     14209 &      1309 &  207.04 \\
    000778 &     90936 &      1022 &   37.38 &     43576 &       532 &   77.06 &     47360 &       426 &   71.36 \\
 \hline \hline
\end{tabular}
\end{table}

\section{Empirical probability distributions}
\label{S3:dist}

\subsection{Empirical distributions}
\label{S3.1:scaling}

\begin{figure}[htb]
\centering
\includegraphics[width=5cm]{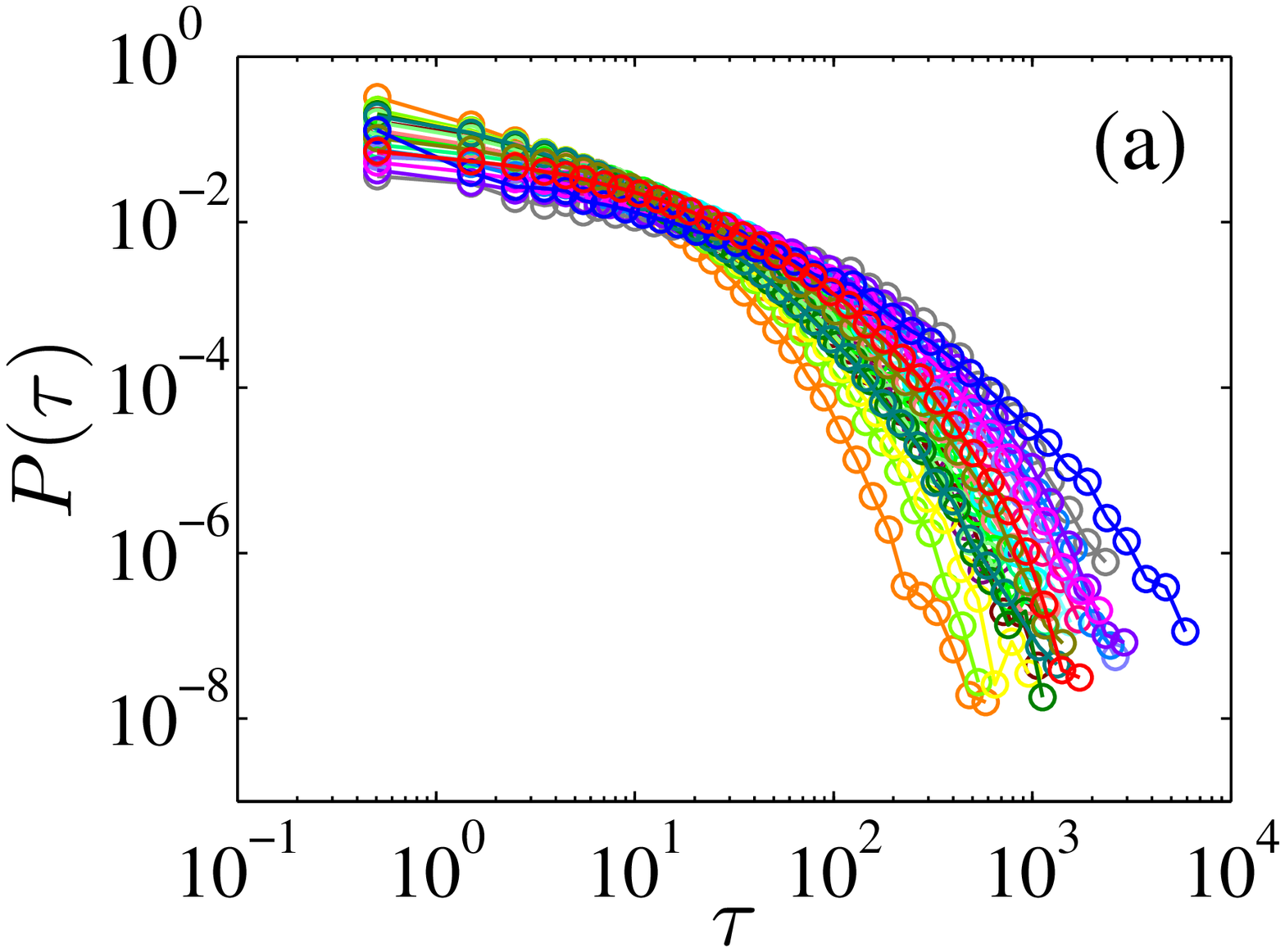}
\includegraphics[width=5cm]{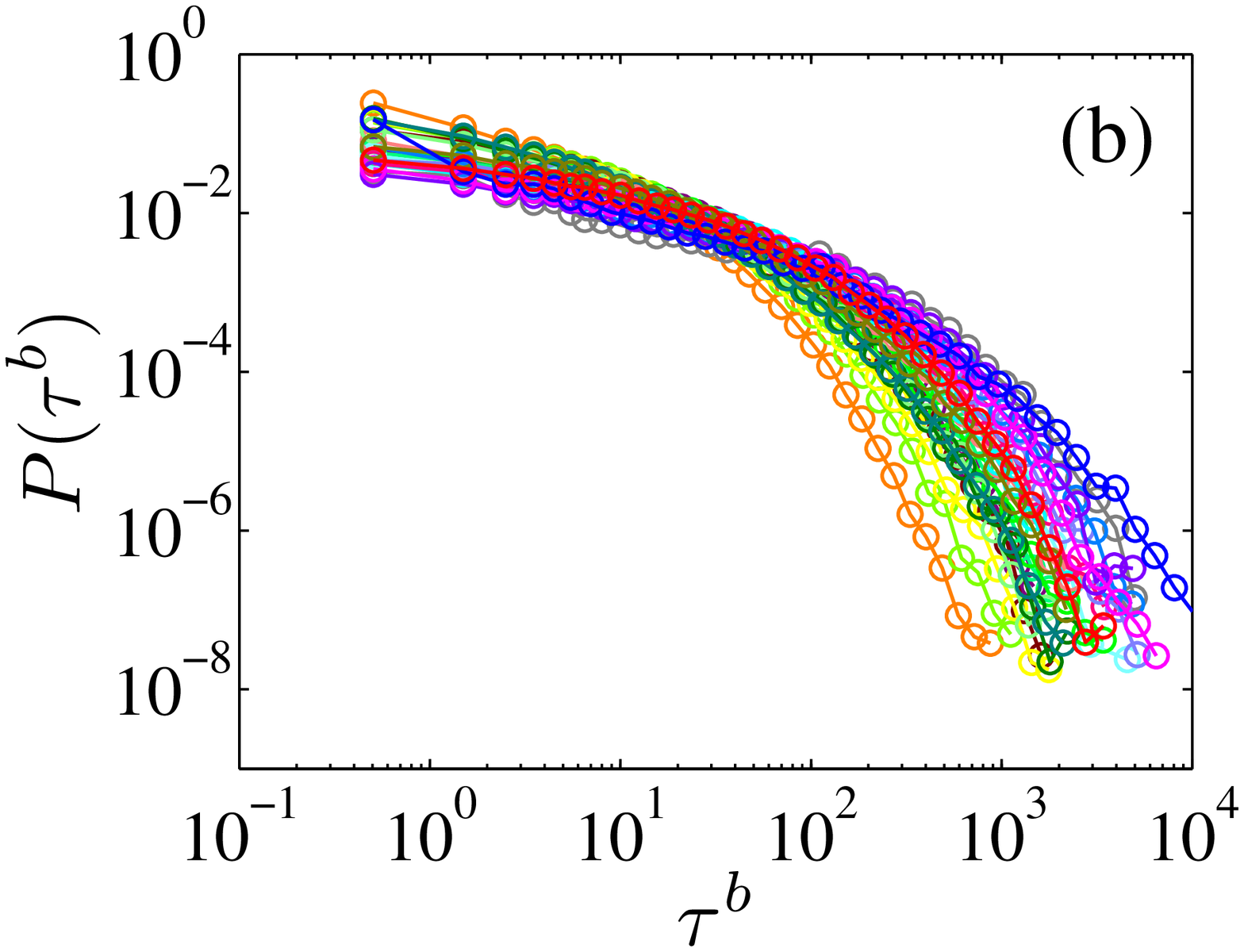}
\includegraphics[width=5cm]{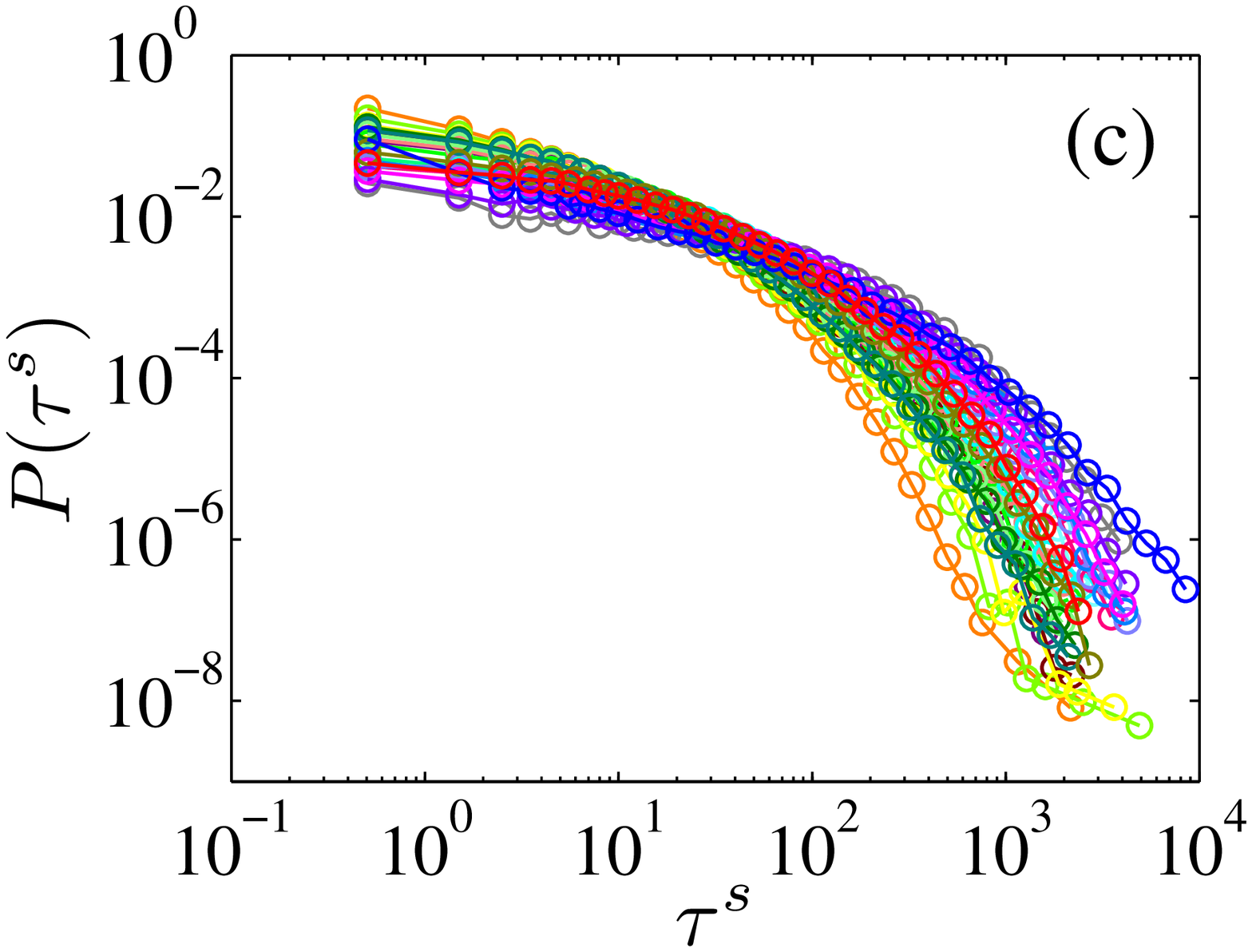}
\includegraphics[width=5cm]{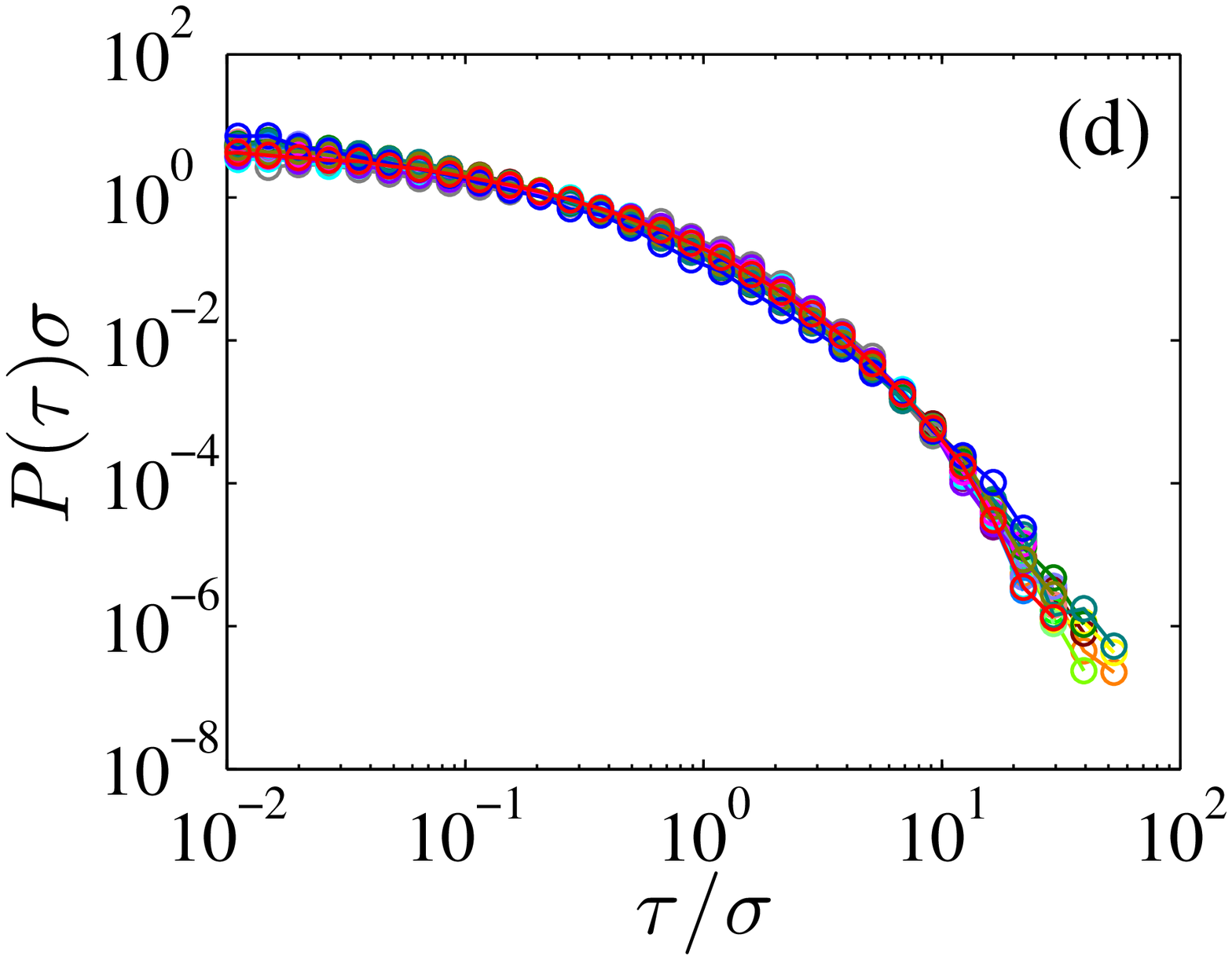}
\includegraphics[width=5cm]{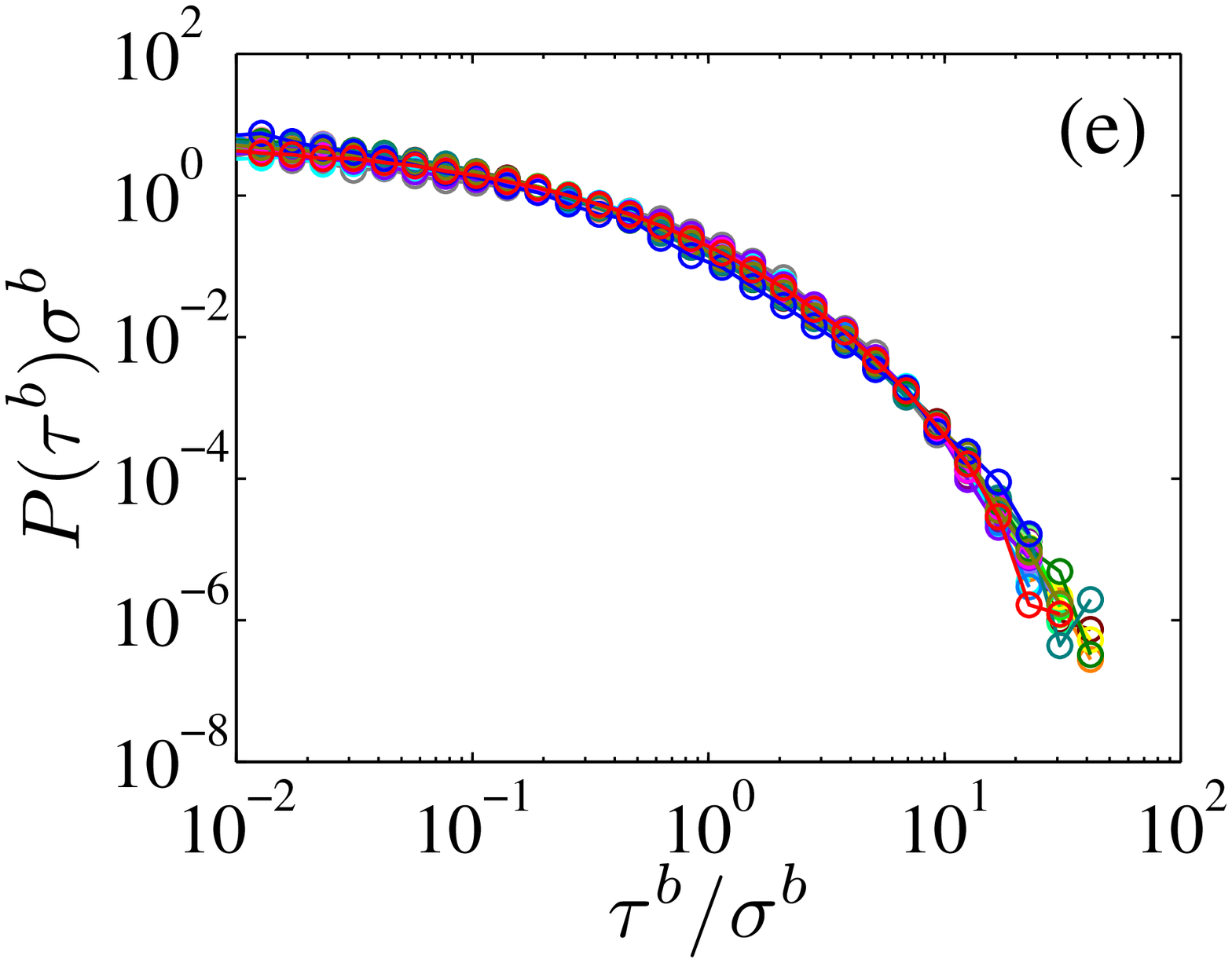}
\includegraphics[width=5cm]{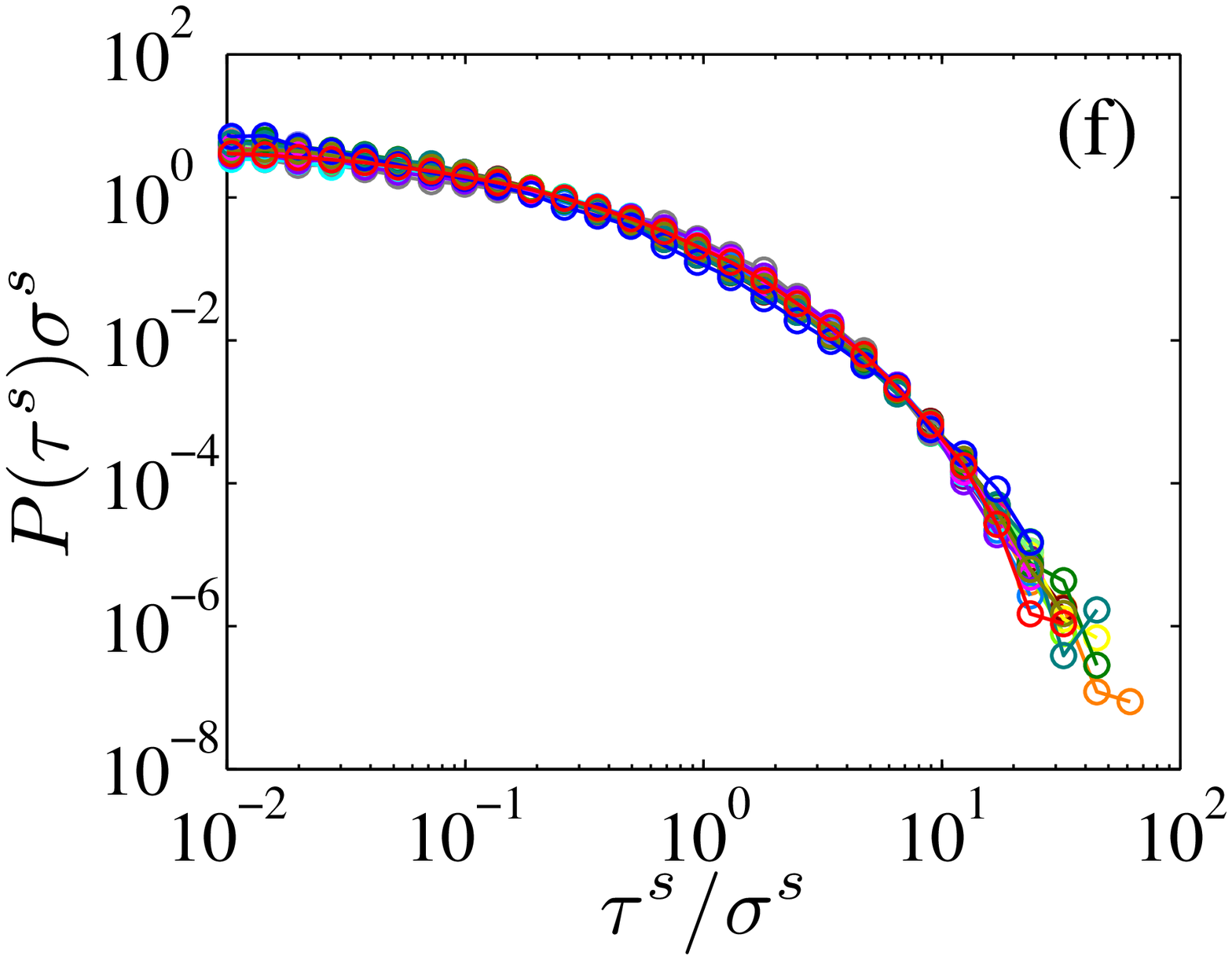}
\includegraphics[width=5cm]{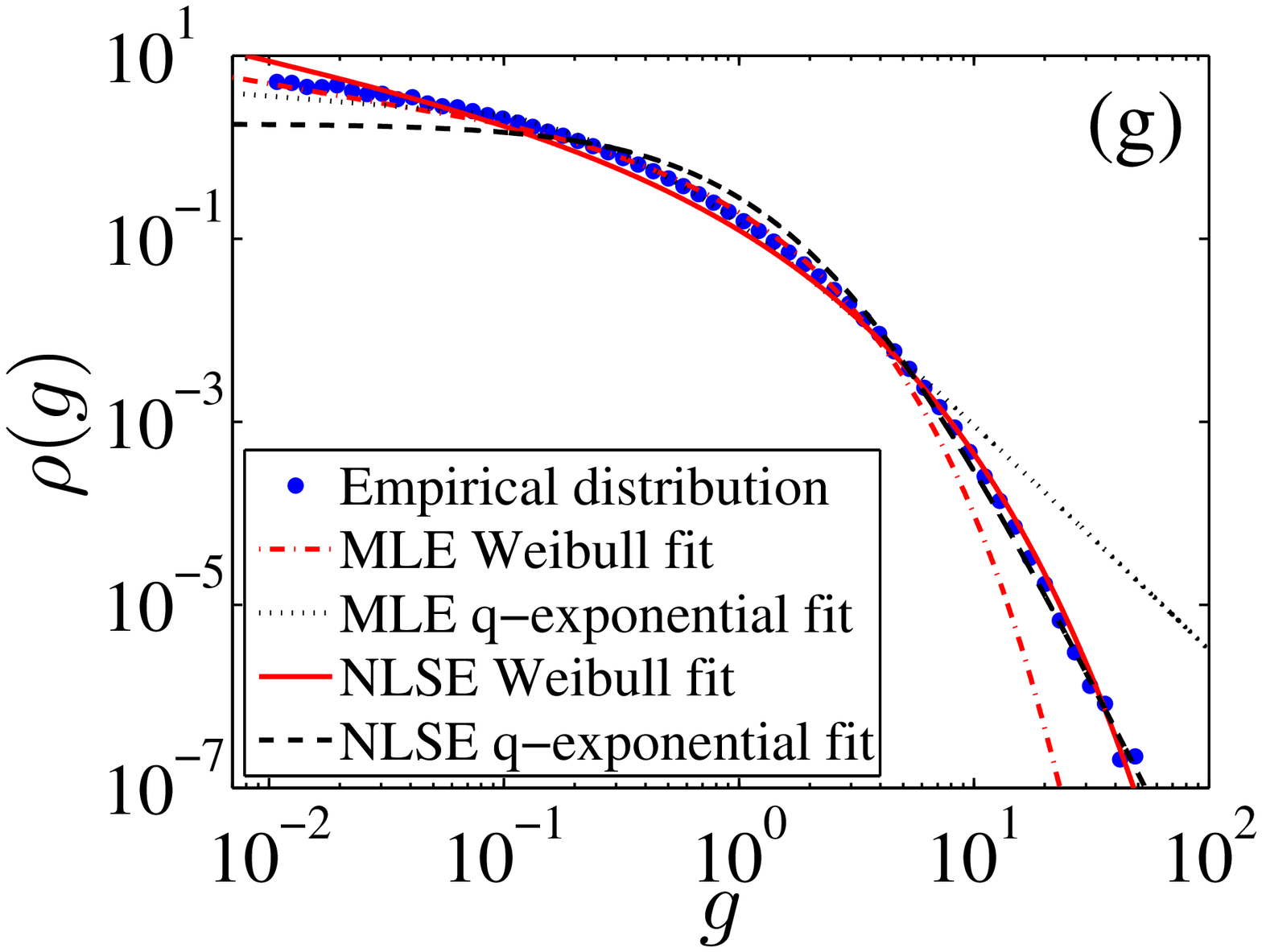}
\includegraphics[width=5cm]{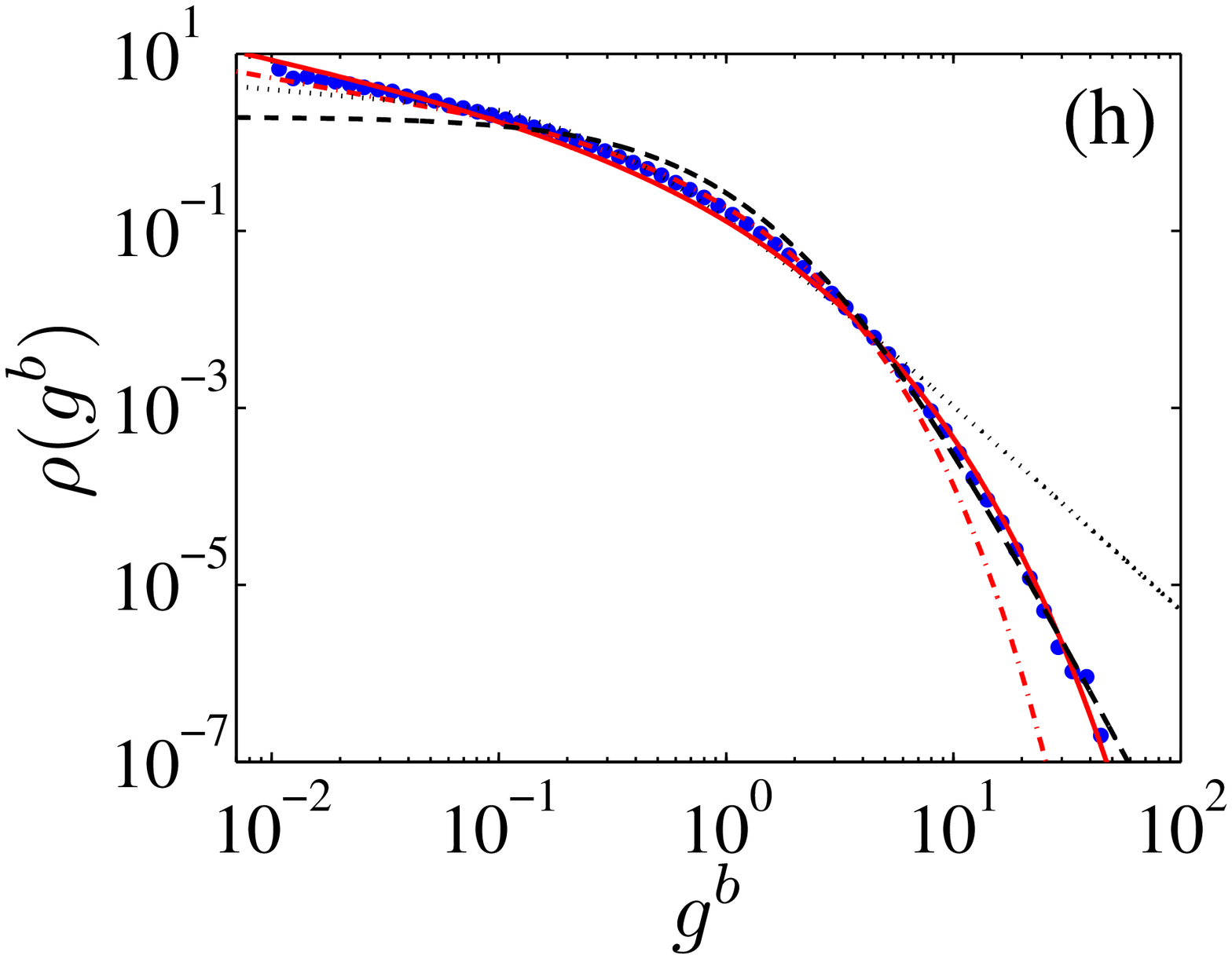}
\includegraphics[width=5cm]{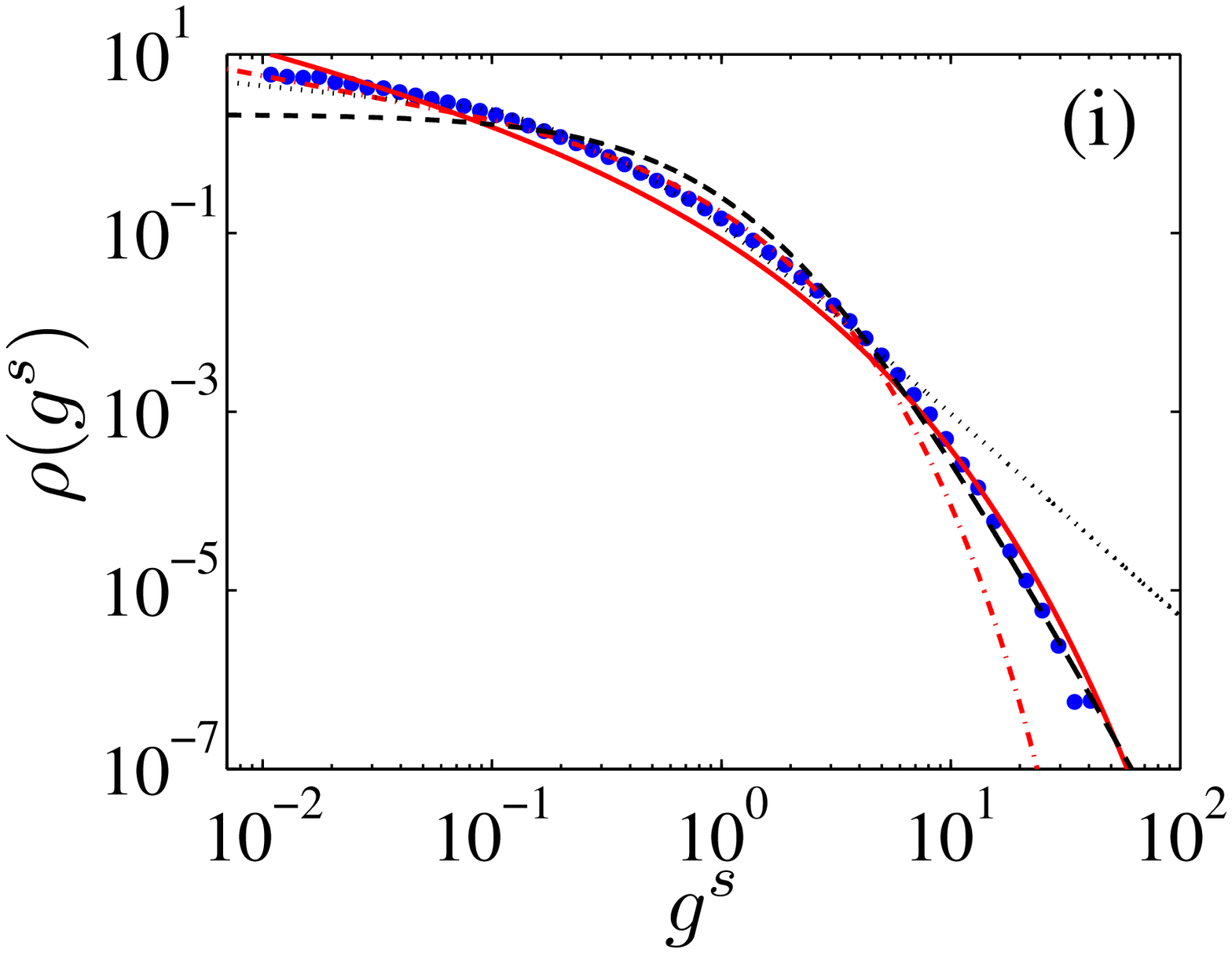}
\caption{\label{Fig:Cancel:PDF} (Color online) Empirical probability distributions of the inter-cancelation durations for all limit orders, buy limit orders and sell limit orders. (a-c) The empirical probability density of inter-cancelation durations for all cancelations, buyer-initialed cancelations and seller-initialed cancelations for the 22 stocks. (d-f) Scaling in distributions of the rescaled inter-cancelation durations. (g-i) Fitting the probability density of the rescaled inter-cancelation durations of the ensemble of all the 22 stocks to the Weibull and the $q$-exponential distributions using the maximum likelihood estimatoion and the nonlinear least-squares regression.}
\end{figure}

In this section, we study the empirical probability distributions of the inter-cancelation durations for
individual stocks. The three types of the empirical distributions of the inter-cancelation durations $\tau$,$\tau^b$ and $\tau^s$ for the 22 individual stocks are plotted in Fig.~\ref{Fig:Cancel:PDF}(a-c). A very important feature is that all these distributions are not exponential. In addition, it is not evident that these distributions have power-law tails by eye-balling. A horizontal comparison of the three plots reveals that the the three distributions $P(\tau)$, $P(\tau^b)$, and $P(\tau^s)$ are very similar.

Similar to the situation of intertrade durations \cite{Ivanov-Yuen-Podobnik-Lee-2004-PRE,Poloti-Scalas-2008-PA,Jiang-Chen-Zhou-2008-PA}, we conjecture that the inter-cancelation duration distributions for individual stocks are Weibulls,
\begin{equation}
 P_{\rm{WBL}}(\tau)=\beta\alpha^{-\beta}\tau^{\beta-1}e^{-(\frac{\tau}{\alpha})^\beta}. \label{Eq:weibull_pdf}
\end{equation}
Following Refs.~\cite{Poloti-Scalas-2008-PA,Jiang-Chen-Zhou-2008-PA}, we also try to model the distributions using $q$-exponentials
\begin{equation}
 P_{\rm{qE}}(\tau)=\frac{1}{\mu}\left[1+(1-q)(-\frac{\tau}{\mu})\right]^{\frac{q}{1-q}}. \label{Eq:qexponential_pdf}
\end{equation}
We have fitted each curve in Fig.~\ref{Fig:Cancel:PDF}(a-c) using the above two equations based on the maximum likelihood estimation (MLE) and the nonlinear least-squares estimation (NLSE). The estimated parameters for the curves in Fig.~\ref{Fig:Cancel:PDF}(a) are depicted in Table \ref{TB:Fit22PDFs}. The r.m.s. of the fitting residuals $\chi_{\rm{WBL}}$ of the Weibull distribution is less than $\chi_{\rm{qE}}$ of the $q$-exponential distribution for 18 stocks when we use the maximum likelihood estimation. When we use the nonlinear least-squares regression, only one stock has $\chi_{\rm{WBL}}> \chi_{\rm{qE}}$. We thus conclude that the inter-cancelation durations are better modeled by the Weibull distribution. For the waiting times between successive cancelations of buy orders and sell orders, we obtain very similar results and the Weibull outperforms the $q$-exponential as well.

\begin{table}[htp]
 \centering
 \caption{Estimated parameters of the Weibull distribution (\ref{Eq:weibull_pdf}) and the $q$-exponential distribution (\ref{Eq:qexponential_pdf}) based on the maximum likelihood estimation (MLE) and the nonlinear least-squares estimation (NLSE), respectively. The columns $\chi_{\rm{WBL}}$ and
$\chi_{\rm{qE}}$ are the r.m.s. values of fit residuals.}
\medskip
\label{TB:Fit22PDFs} \centering
\begin{tabular}{lrrrrrcrrrrrc}
 \hline \hline
  \multicolumn{1}{c}{Code} &
  \multicolumn{6}{c}{\begin{tabular}{rrrrrr} \multicolumn{6}{c}{~~~~MLE}~~~~ \\ \hline $\alpha$~~~ & $\beta$~~~ & $\chi_{\rm{WBL}}$~~~ & $\mu$~~~ & $q$~~~ & $\chi_{\rm{qE}}$
  \end{tabular}} &
  \multicolumn{6}{c}{\begin{tabular}{rrrrrr} \multicolumn{6}{c}{~~~~NLSE}~~~~ \\ \hline $\alpha$~~~ & $\beta$~~~ & $\chi_{\rm{WBL}}$~~~ & $\mu$~~~ & $q$~~~ & $\chi_{\rm{qE}}$
  \end{tabular}}\\
  \hline
    000001 &    0.41 &    0.67 &    0.42 &    0.24 &    1.67 &    0.51 &    0.18 &    0.47 &    0.85 &    0.49 &    1.26 &    1.14\\
    000002 &    0.47 &    0.71 &    0.15 &    0.30 &    1.57 &    0.30 &    0.34 &    0.57 &    0.60 &    0.50 &    1.27 &    0.77\\
    000009 &    0.42 &    0.69 &    0.42 &    0.26 &    1.62 &    0.54 &    0.31 &    0.55 &    0.54 &    0.53 &    1.24 &    1.14\\
    000012 &    0.35 &    0.66 &    0.31 &    0.19 &    1.71 &    0.25 &    0.22 &    0.49 &    0.66 &    0.43 &    1.30 &    1.14\\
    000016 &    0.48 &    0.71 &    0.17 &    0.32 &    1.54 &    0.27 &    0.32 &    0.56 &    0.74 &    0.52 &    1.25 &    0.71\\
    000021 &    0.39 &    0.68 &    0.29 &    0.23 &    1.65 &    0.39 &    0.18 &    0.47 &    0.81 &    0.46 &    1.28 &    1.05\\
    000024 &    0.44 &    0.69 &    0.16 &    0.28 &    1.60 &    0.33 &    0.29 &    0.54 &    0.62 &    0.48 &    1.28 &    0.86\\
    000027 &    0.42 &    0.69 &    0.24 &    0.26 &    1.61 &    0.30 &    0.27 &    0.53 &    0.76 &    0.46 &    1.29 &    0.82\\
    000063 &    0.36 &    0.64 &    0.39 &    0.18 &    1.81 &    0.42 &    0.26 &    0.52 &    0.37 &    0.44 &    1.31 &    1.34\\
    000066 &    0.38 &    0.67 &    0.29 &    0.22 &    1.67 &    0.34 &    0.24 &    0.50 &    0.71 &    0.41 &    1.33 &    0.96\\
    000088 &    0.56 &    0.72 &    0.40 &    0.41 &    1.45 &    0.62 &    0.44 &    0.63 &    0.51 &    0.55 &    1.25 &    0.79\\
    000089 &    0.51 &    0.73 &    0.15 &    0.35 &    1.49 &    0.24 &    0.40 &    0.61 &    0.59 &    0.55 &    1.24 &    0.63\\
    000429 &    0.42 &    0.68 &    0.19 &    0.25 &    1.66 &    0.31 &    0.31 &    0.55 &    0.49 &    0.48 &    1.28 &    0.96\\
    000488 &    0.41 &    0.67 &    0.43 &    0.24 &    1.68 &    0.55 &    0.28 &    0.53 &    0.47 &    0.46 &    1.30 &    1.19\\
    000539 &    0.39 &    0.63 &    0.29 &    0.21 &    1.79 &    0.43 &    0.35 &    0.58 &    0.30 &    0.51 &    1.27 &    1.20\\
    000541 &    0.51 &    0.71 &    0.14 &    0.34 &    1.52 &    0.36 &    0.36 &    0.58 &    0.60 &    0.54 &    1.24 &    0.73\\
    000550 &    0.33 &    0.64 &    0.34 &    0.17 &    1.78 &    0.30 &    0.19 &    0.47 &    0.58 &    0.42 &    1.31 &    1.29\\
    000581 &    0.44 &    0.69 &    0.23 &    0.27 &    1.62 &    0.39 &    0.34 &    0.57 &    0.46 &    0.51 &    1.26 &    0.94\\
    000625 &    0.34 &    0.65 &    0.34 &    0.18 &    1.73 &    0.24 &    0.15 &    0.44 &    0.90 &    0.40 &    1.33 &    1.08\\
    000709 &    0.39 &    0.67 &    0.22 &    0.23 &    1.66 &    0.20 &    0.26 &    0.52 &    0.67 &    0.47 &    1.28 &    0.98\\
    000720 &    0.23 &    0.52 &    0.26 &    0.08 &    2.30 &    1.06 &    0.17 &    0.45 &    0.43 &    0.30 &    1.45 &    1.13\\
    000778 &    0.45 &    0.70 &    0.18 &    0.29 &    1.59 &    0.25 &    0.33 &    0.56 &    0.62 &    0.55 &    1.23 &    0.86\\
  \hline \hline
\end{tabular}
\end{table}

Table \ref{TB:Fit22PDFs} illustrates that the estimated values of each parameter for different stocks are close to the mean, especially for $\beta$. In addition, the shapes of the empirical distributions shown in Fig.~\ref{Fig:Cancel:PDF}(a-c) are very similar to those of the intertrade durations \cite{Ivanov-Yuen-Podobnik-Lee-2004-PRE,Poloti-Scalas-2008-PA,Jiang-Chen-Zhou-2008-PA}. To the best of our knowledge, a scaling behavior in the rescaled intertrade duration distributions was first reported by Ch. Ivanov et al. \cite{Ivanov-Yuen-Podobnik-Lee-2004-PRE} with further evidence provided by Jiang et al. \cite{Jiang-Chen-Zhou-2008-PA}. In order to check if the distributions of inter-cancelation durations also have a scaling behavior, we apply the following rescaling scheme to each stock
\begin{equation}
 \tau\rightarrow\tau/\sigma,~P\rightarrow \sigma P,
\label{Eq:Rescale}
\end{equation}
where $\sigma$ is the standard deviation of the durations of a given stock. Figure \ref{Fig:Cancel:PDF}(d-f) plots the rescaled probability $\sigma P(\tau)$ as a function of the rescaled duration $\tau/\sigma$ for the three types of inter-cancelation durations. We find in each plot that all the 22 curves collapse onto a single curve, which implies a scaling form
\begin{equation}
 \sigma P(\tau) = \rho(\tau/\sigma),
\label{Eq:scaling}
\end{equation}
where $\rho$ is the scaling function.

Since the rescaled distributions of inter-cancelation durations for different stocks show a nice scaling, it enables us to treat all the rescaled inter-cancelation durations from different stocks as an ensemble to gain better statistics. We use $g$, $g^b$ and $g^s$ to denote the variables $\tau/\sigma$, $\tau^b/\sigma^b$ and $\tau^s/\sigma^s$ from 22 different stocks, respectively. Figure \ref{Fig:Cancel:PDF}(g-i) illustrates the three empirical probability densities of the three types of rescaled intercancelation durations. Again, we use the Weibull and the $q$-exponential to model each distribution $\rho(\tau/\sigma)$ based on the maximum likelihood estimation and the nonlinear least-squares estimation. The fitted curves are also drawn in Fig.~\ref{Fig:Cancel:PDF}(g-i), and the estimated parameters are listed in Table \ref{Tb:Fit3}. We find that $\chi_{\rm{WBL}} < \chi_{\rm{qE}}$ for all the cases except for the inter-cancelation durations of sell limit orders $g^s$ based on the maximum likelihood estimation where $\chi_{\rm{WBL}}$ is slightly greater than $\chi_{\rm{qE}}$. Therefore, the Weibull distribution is a better model for the three scaling functions. We also observe that the three $\beta$ values are close to each other for either the maximum likelihood estimation or the nonlinear least-squares regression. It is interesting to note that the $\beta$ values are also close to those for the intertrade durations \cite{Jiang-Chen-Zhou-2008-PA}.

\begin{table}[htp]
 \centering
 \caption{Estimated values of parameters ($\alpha$, $\beta$, $q$, $\mu$) by means of MLE and NLSE. $\chi_{\rm{WBL}}$ and $\chi_{\rm{qE}}$ stand for the r.m.s. of fit residuals.}
 \medskip
 \label{Tb:Fit3}
 \begin{tabular}{cccccccccccccccccccccc}
  \hline \hline
%  \multicolumn{1}{c}{Duration} &
  & \multicolumn{6}{c}{MLE} && \multicolumn{6}{c}{NLSE} \\
  \cline{2-7} \cline{9-14}
             &$\alpha$&$\beta$&$\chi_{\rm{WBL}}$ & $\mu$ &$q$ &$\chi_{\rm{qE}}$ && $\alpha$&$\beta$&$\chi_{\rm{WBL}}$ & $\mu$ &$q$ &$\chi_{\rm{qE}}$ \\
  \hline
    $g$~~~   &  0.40 &  0.67 &  0.21 &  0.23 &  1.67 &  0.30 &&   0.21 &    0.49 &    0.70 &    0.55 &    1.23 &    1.13 \\
    $g^b$~~  &  0.38 &  0.65 &  0.27 &  0.21 &  1.75 &  0.43 &&   0.23 &    0.50 &    0.45 &    0.51 &    1.25 &    1.28 \\
    $g^s$~~  &  0.34 &  0.64 &  0.26 &  0.18 &  1.78 &  0.24 &&   0.11 &    0.42 &    0.84 &    0.47 &    1.27 &    1.26 \\
  \hline \hline
\end{tabular}
\end{table}

We now focus on the curves in Fig.~\ref{Fig:Cancel:PDF}(g) fitted using the Weibull distribution. It is found that the maximum likelihood estimation curve fits the bulk of the distribution very well in the range $[10^{-2},5]$, which accounts for 99.1\% of the sample. However, it is unable to capture the large durations. In contrast, the nonlinear least-squares regression curve fits the tail quite well but deviates markedly the bulk of the distributions.

\subsection{Conditional distributions of inter-cancelation durations}
\label{S2:Cancel:CondPDF}

We now investigate the conditional distribution of rescaled inter-cancelation durations on the value of its preceding duration. All the rescaled durations for different stocks constitute an ensemble set $Q$. We sort the set $Q$ in an increasing order and partition it into eight non-overlapping groups of the same size:
\begin{equation}
 Q=\bigcup_{i=1}^{8}Q_i~, \label{Eq:conditional_dist}
\end{equation}
where $Q_i \cap Q_j = \phi$ when $i \neq j$, and $g_i<g_j$ when $g_i\in Q_i$, $g_j\in Q_j$, $i<j$. We then estimate the empirical conditional probability distribution $P(g|g_0\in Q_i)\triangleq P(g(t)|g(t-1)\in Q_i)$, which is the probability density of the rescaled inter-cancelation durations $g(t)$ whose preceding value $g_0=g(t-1)$ belongs to $Q_i$.

\begin{figure}[htb]
\centering
\includegraphics[width=7.5cm]{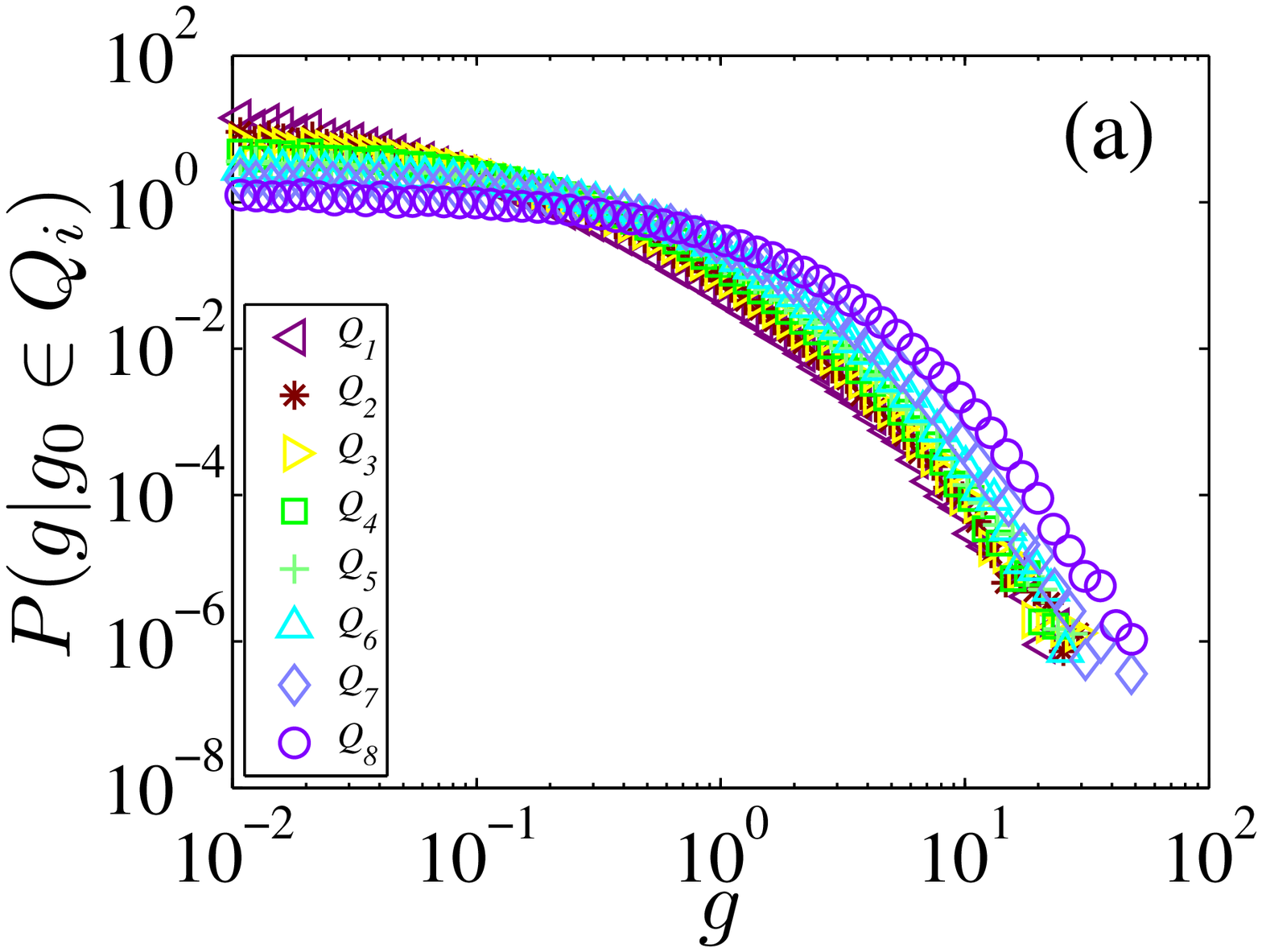}
\includegraphics[width=7.5cm]{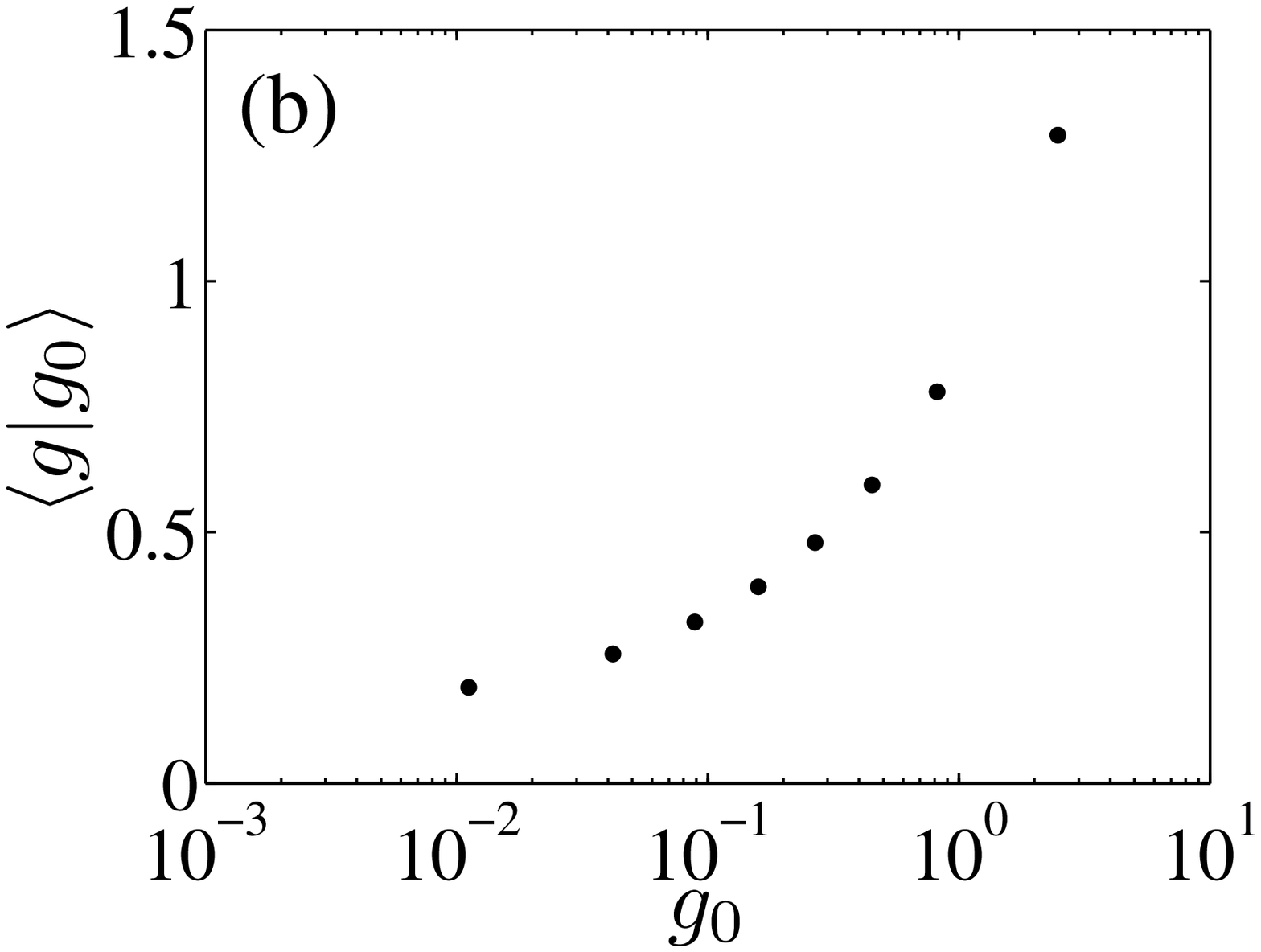}
\caption{\label{Fig:Cancel:CondPDF} (Color online.) (a) Conditional probability density $P(g|g_0\in Q_i)$. (b) Conditional mean inter-cancelation durations $\langle g|g_0\rangle$ with respect to $g_0=\langle{g:g\in Q_i}\rangle$.}
\end{figure}

The eight empirical conditional PDFs are plotted in Fig.~\ref{Fig:Cancel:CondPDF}(a). Assuming that $i>j$, we note that $P(g|g_0\in Q_i)<P(g|g_0\in Q_j)$ for small $g$ and $P(g|g_0\in Q_i)>P(g|g_0\in Q_j)$ for small $g$. In other words, large durations tend to follow large durations and small durations are prone to follow small durations. Figure \ref{Fig:Cancel:CondPDF}(b) shows the dependence of the conditional mean duration $\langle g|g_0\rangle$ with respect to $g_0$. It is shown that the conditional mean duration increases with $g_0$, which is consistent with the outcome of Fig.~\ref{Fig:Cancel:CondPDF}(a). This phenomenon indicates that there is short-term memory in the inter-cancelation durations.

We recall that the conditional distributions of the intertrade durations with respect to different preceding durations almost collapse onto a single curve and the conditional mean of intertrade durations does not changes with the preceding durations \cite{Jiang-Chen-Zhou-2008-PA}. Therefore, although both the intertrade durations and the inter-cancelation durations exhibit scaling in the distributions and the scaling functions are both Weibulls with very close exponents, the difference in the behavior of conditional durations unveils that transactions and cancelations follow different dynamic processes.

\section{Memory effects}
\label{S1:Memory}

\subsection{Intraday pattern}
\label{S2:IntradayPatten}

\begin{figure}[htb]
\centering
\includegraphics[width=4cm]{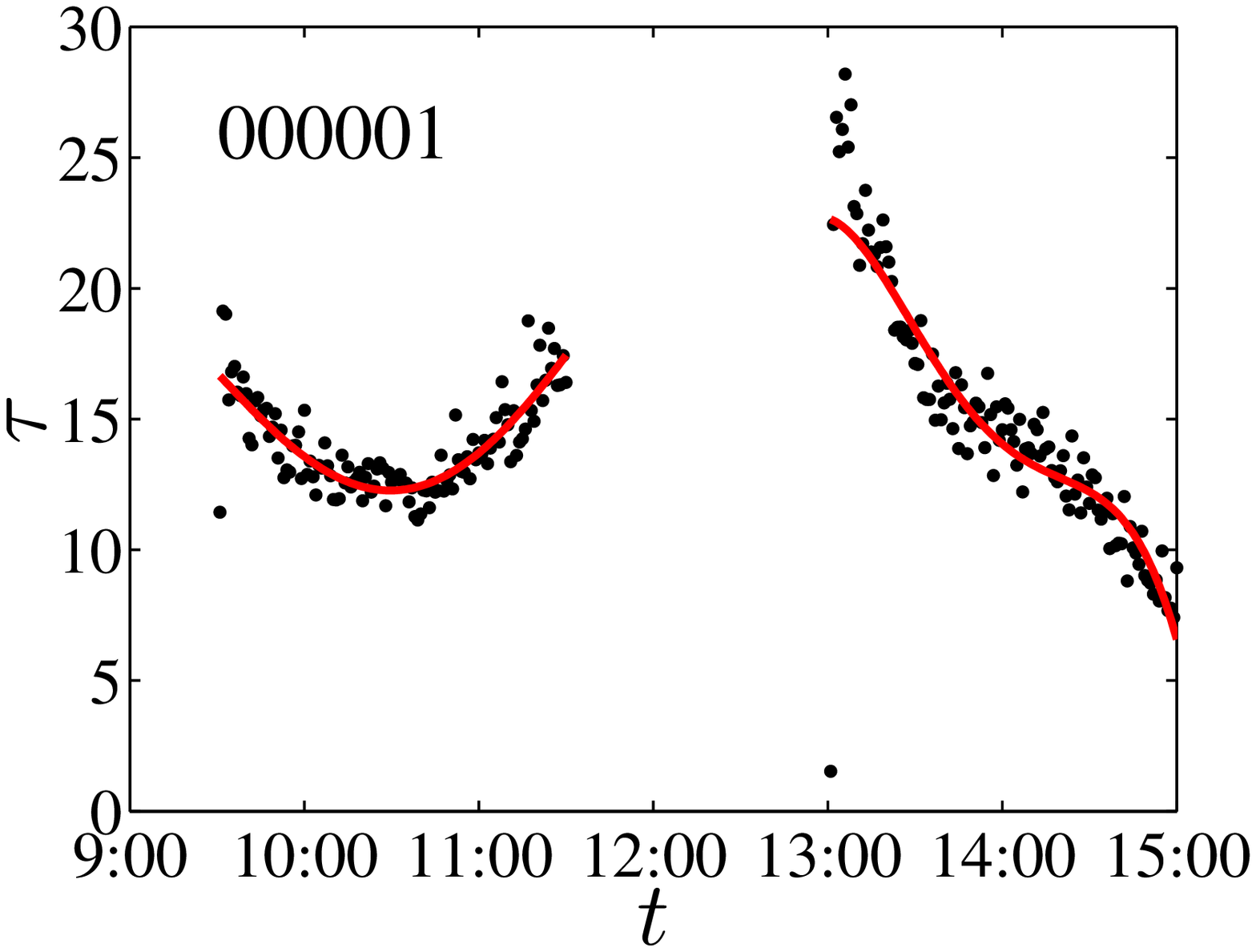}
\includegraphics[width=4cm]{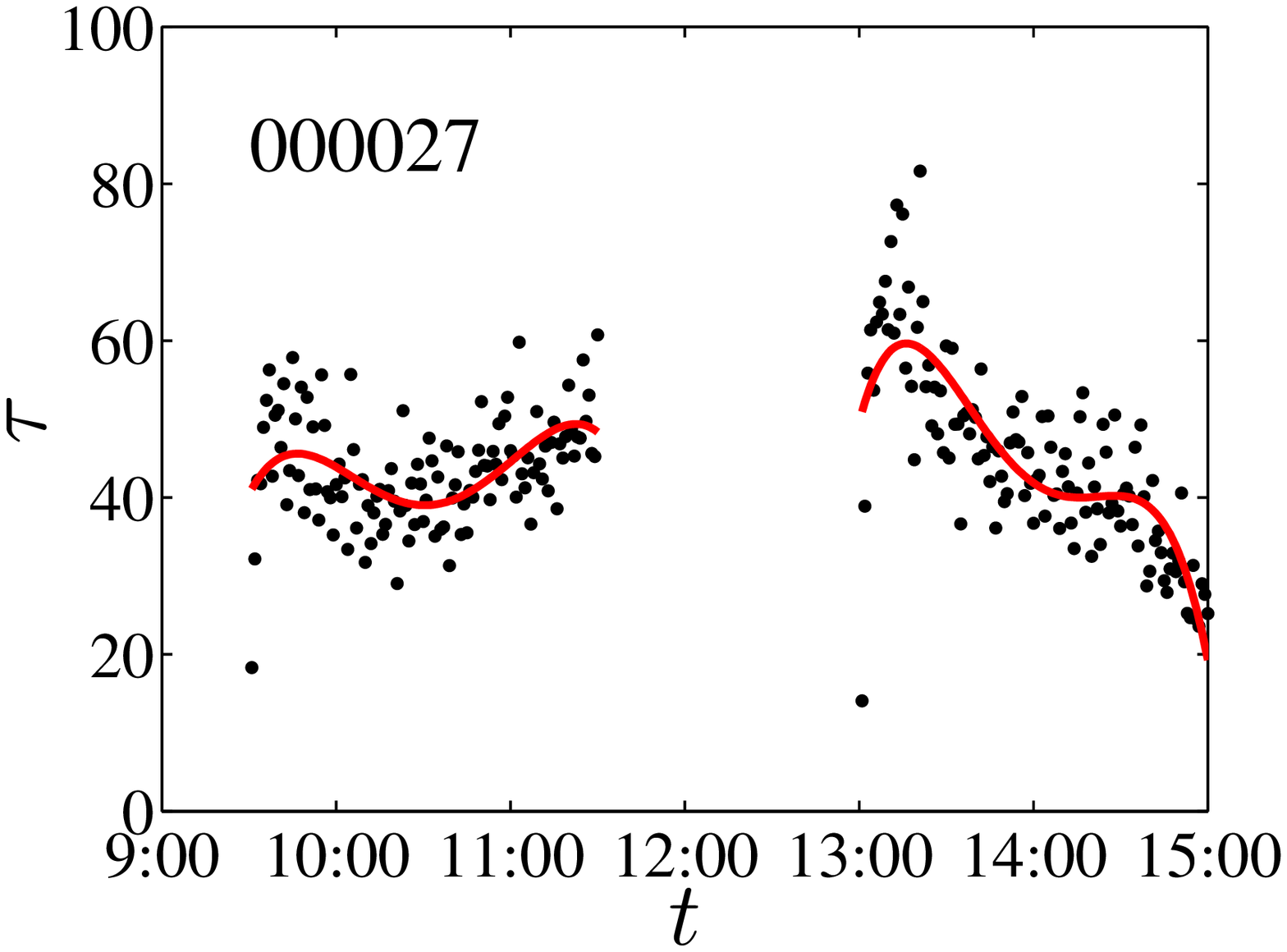}
\includegraphics[width=4cm]{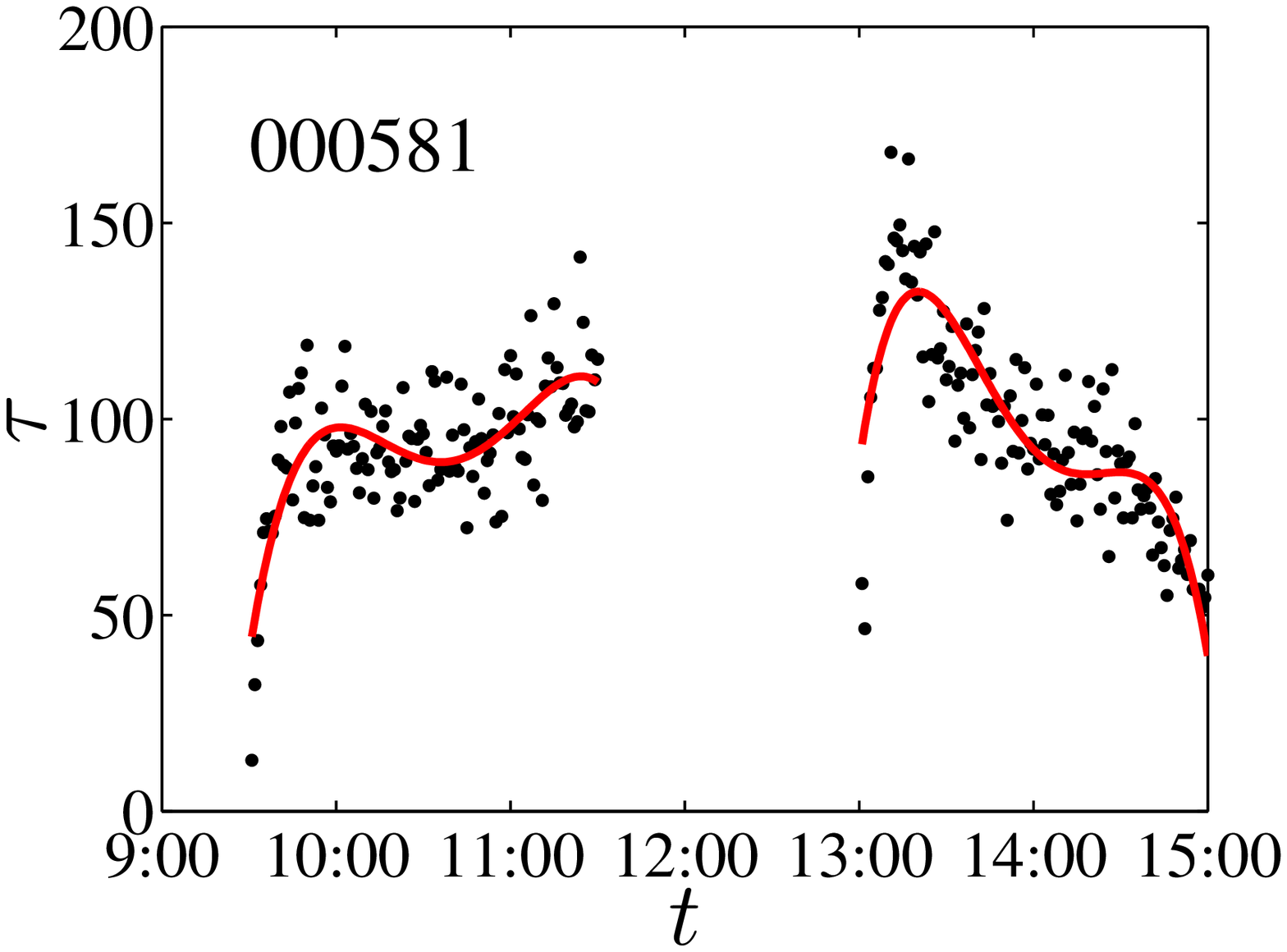}
\includegraphics[width=4cm]{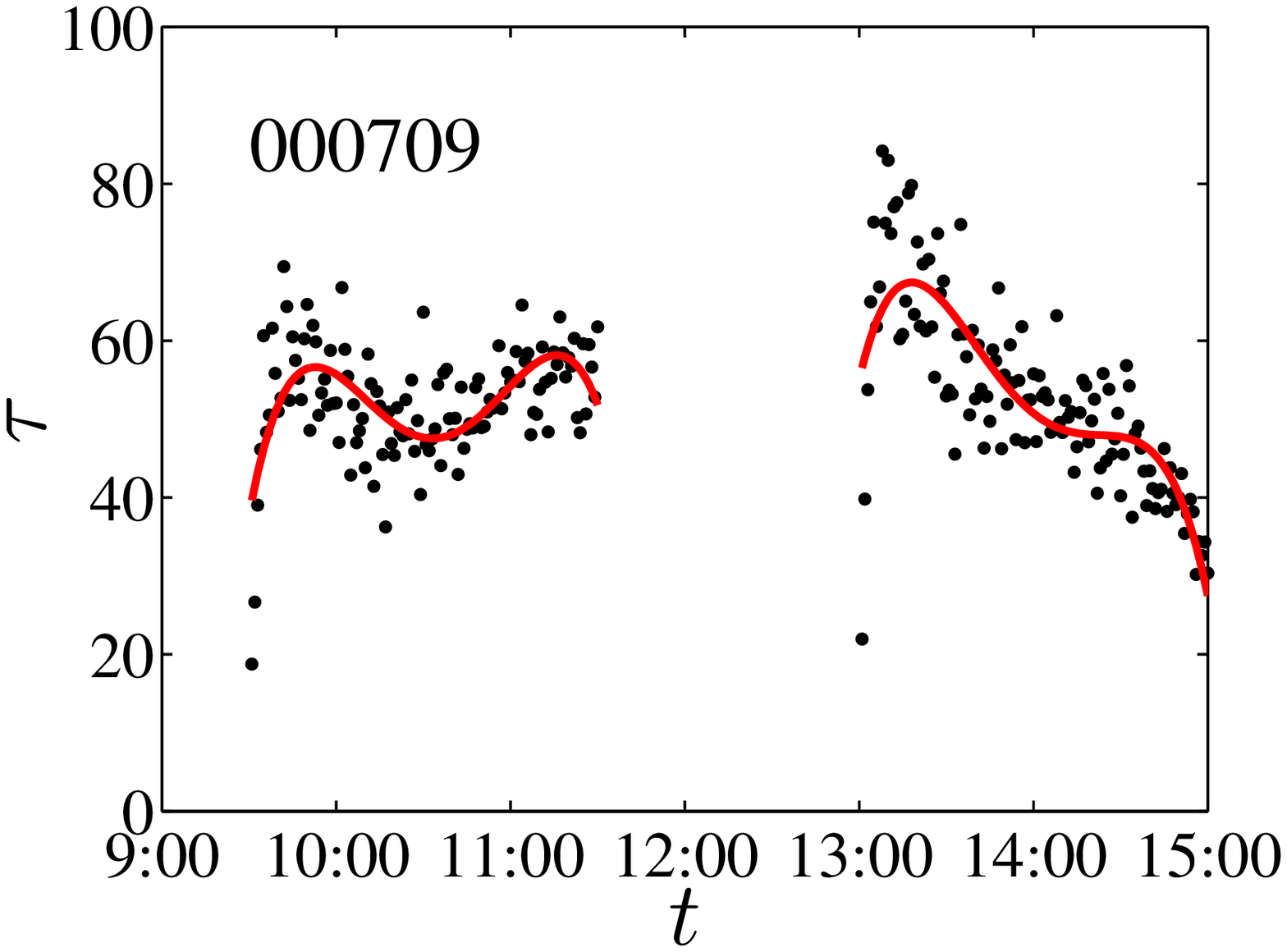}
\caption{\label{Fig:intradaypattern} (Color online) Intraday pattern
of inter-cancelation durations for four different stocks traded on
the SZSE during the calendar year 2003. The dots show the cross
sectional mean values of inter-cancelation durations over the whole
trading year at that minute. The continuous curves are the
polynomial fits to the data. From the curves, a crude inverse
$U$-shaped pattern can be observed.}
\end{figure}

Intraday patterns exist in many high-frequency financial variables in the empirical studies. It is necessary to investigate the intraday patterns in the inter-cancelation durations for each stock. To obtain the intraday patterns of inter-cancelation durations, we segment the continuous double auction of each trading day into 240 successive 1-min intervals. For a given stock, the inter-cancelation durations are then averaged within each trading minute to create a minute-by-minute series as follows,
\begin{equation}
 \tau_{ij}=\frac{1}{N_{ij}}\sum_{k=1}^{N_{ij}}\tau_k, \label{Eq:intra_tau}
\end{equation}
where $N_{ij}$ represents the number of inter-cancelation durations in the $j$-th interval in the $i$-th trading day, $\tau_k$ is the inter-cancelation duration of an order which is canceled in the $j$-th minuite, and $\tau_{ij}$ is the average duration of the $j$-th interval in the $i$-th trading day. The mean inter-cancelation duration at a given minute of a trading day is calculated as follows,
\begin{equation}
 \langle\tau\rangle_{j}=\frac{1}{N_{d}}\sum_{i=1}^{N_{d}}\tau_{ij}~, \label{Eq:intra_tau2}
\end{equation}
where $N_d$ is the number of trading days. Figure \ref{Fig:intradaypattern} plots $\langle\tau\rangle_{j}$ as a function of the intraday time for four randomly chosen stocks (000001, 000027, 000581, 000709). Complex intraday patterns are observed, which share some analogues with those of intertrade durations \cite{Jiang-Chen-Zhou-2009-PA}.

Since the inter-cancelation duration shows an explicit intraday pattern, we define the socalled adjusted inter-cancelation durations by removing the
intraday patterns from the original data,
\begin{equation}
 \tilde{\tau}_t = \tau_t/\langle\tau\rangle_j. \label{Eq:tau}
\end{equation}
We will analyze both the original data and the adjusted data in the following, which allows us to determine if the intraday patterns have influence on the temporal correlations.

\subsection{Long-range dependence}
\label{S4.2:DFA}

To detect the long-range dependence of the inter-cancelation durations, the detrended fluctuation analysis (DFA) is utilized,
which is able to extract long-range power-law correlation in non-stationary time series \cite{Peng-Buldyrev-Havlin-Simons-Stanley-Goldberger-1994-PRE,Kantelhardt-Bunde-Rego-Havlin-Bunde-2001-PA}. The DFA procedure consists of following steps. For a given inter-cancelation duration series $\{\tau_i|i = 1, 2, ..., N\}$, the cumulative summation series $y_i$ should be first calculated as follows,
\begin{equation}
 y_i=\sum_{j=1}^{i}\tau_{j},~~~~i=1,2,...,N. \label{Eq:yi}
\end{equation}
Then we use $N_s$ disjoint intervals with the same size $s$ to cover the series $y$. Since the length of the series $N$ need not be a multiple of the size of the interval $s$, the whole series $y_i$ may not be completely covered by $N_s$ intervals. For compensating the remain part, we can use another $N_s$ intervals to cover the series from the other end of the series. In each interval, a polynomial is used to calculate the local trend function $\tilde{y}$ by least-squares regressions. In this section, linear functions are used in the fitting procedure. The local detrended fluctuation function $r_k(s)$ in the $k$-th interval is defined as the r.m.s. of the fitting residuals:
\begin{equation}
 [r_k(s)]^2=\frac{1}{s}\sum_{i\in I_k}[y_i-\tilde{y}_i]^2~, \label{Eq:fk}
\end{equation}
where $I_k$ is the $k$-th interval. Thus the overall detrended fluctuation is then estimated as follows
\begin{equation}
 [F_2(s)]^2=\frac{1}{2N_s}\sum_{k=1}^{2N_s}[r_k(s)]^2. \label{Eq:F2}
\end{equation}
To determine the scaling behavior of the fluctuation function, we vary the scale $s$ in the range of $[20,N/4]$ (scale $s>N/4$ is excluded since the detrended fluctuation $F_2(s)$ becomes statistically unreliable). Thus a power-law relationship between the overall fluctuation function $F_2(s)$ and
the interval size $s$ can be expected as follows,
\begin{equation}
 F_2(s)\sim s^H, \label{Eq:hurst}
\end{equation}
where $H$ stands for the DFA scaling exponent. Practically, we can plot $F_2(s)$ as a function of $s$ on double logarithmic scales to measure $H$ by a linear fit.

\begin{figure}[htb]
\centering
\includegraphics[width=4cm]{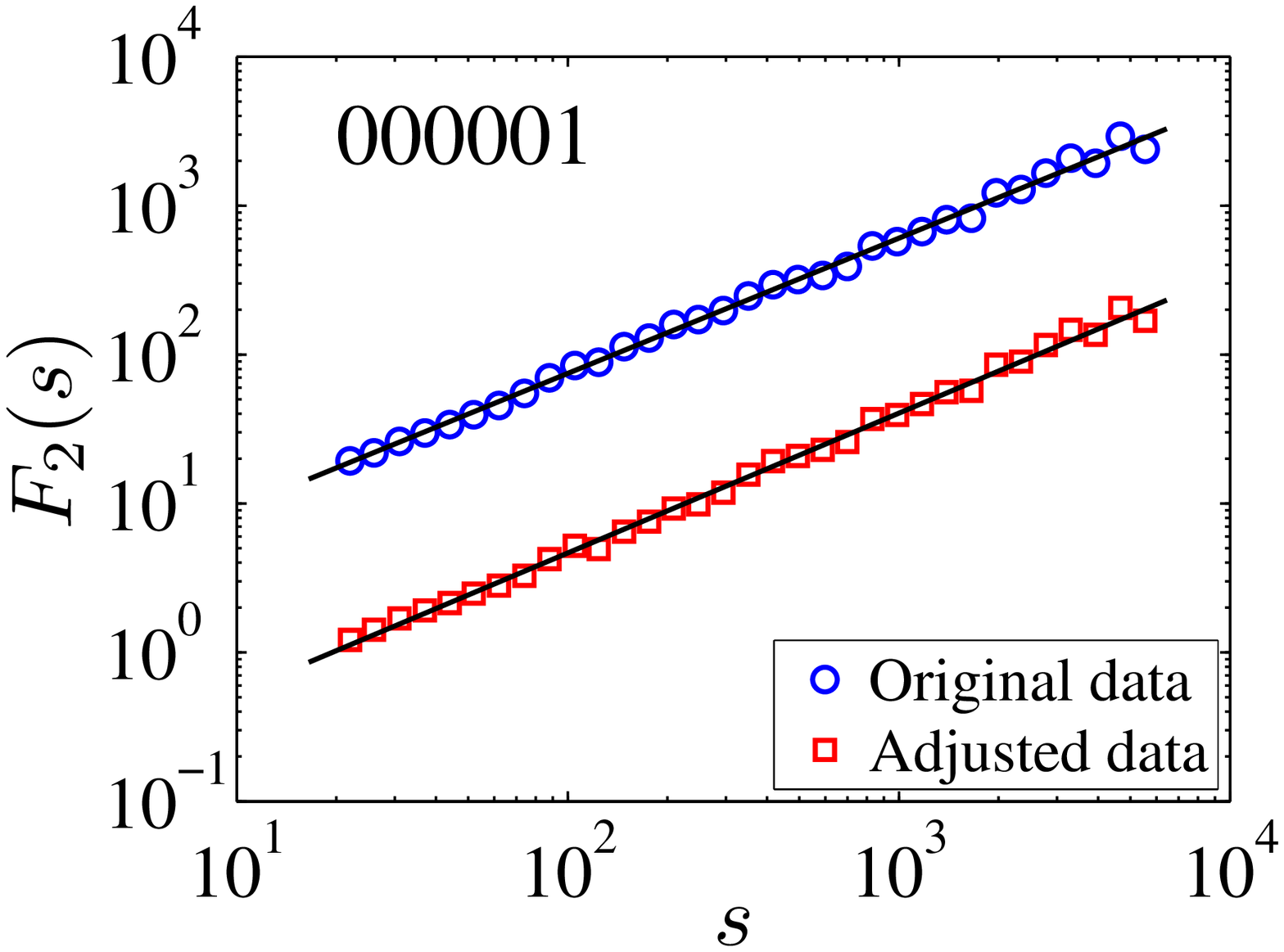}
\includegraphics[width=4cm]{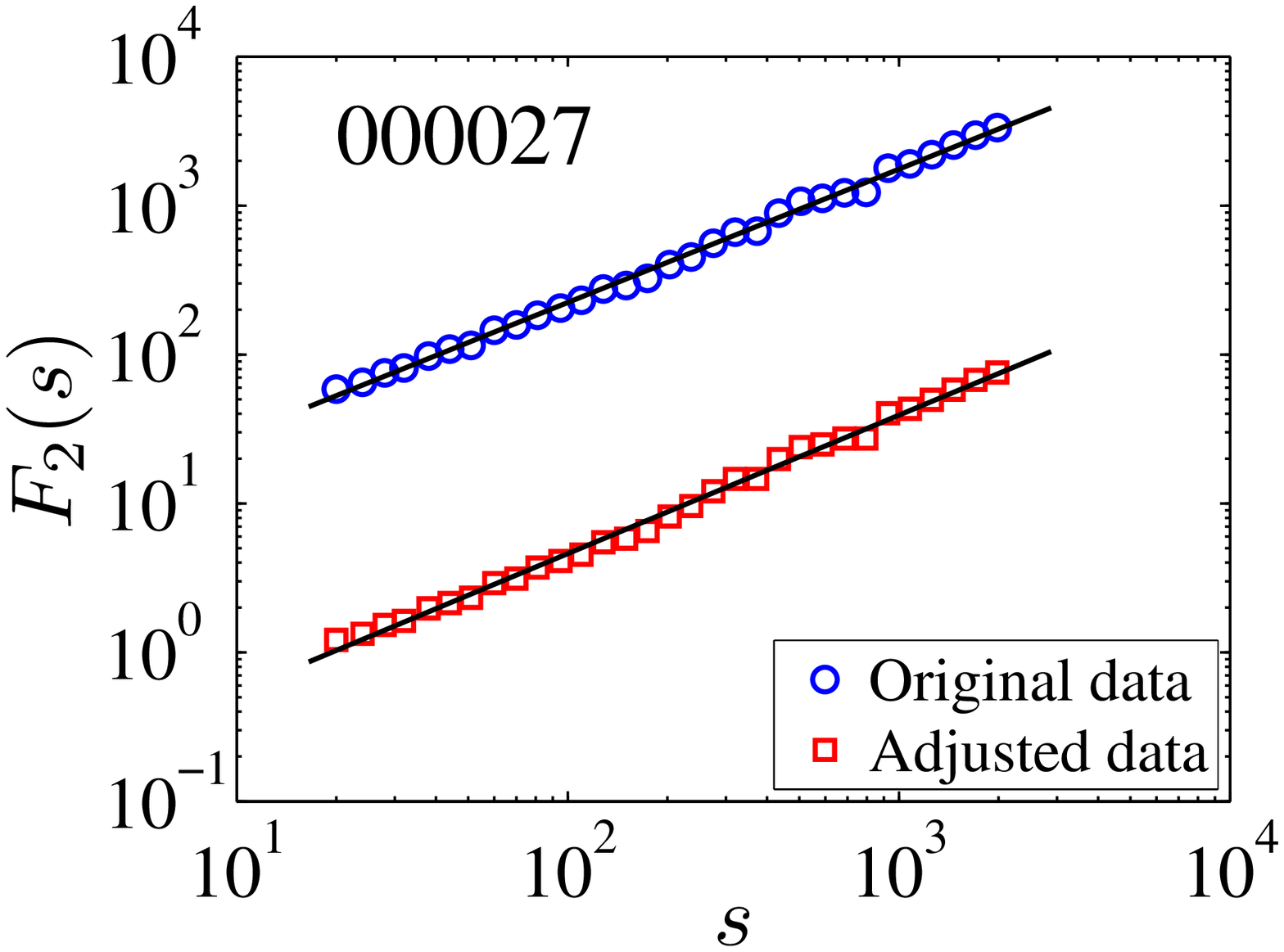}
\includegraphics[width=4cm]{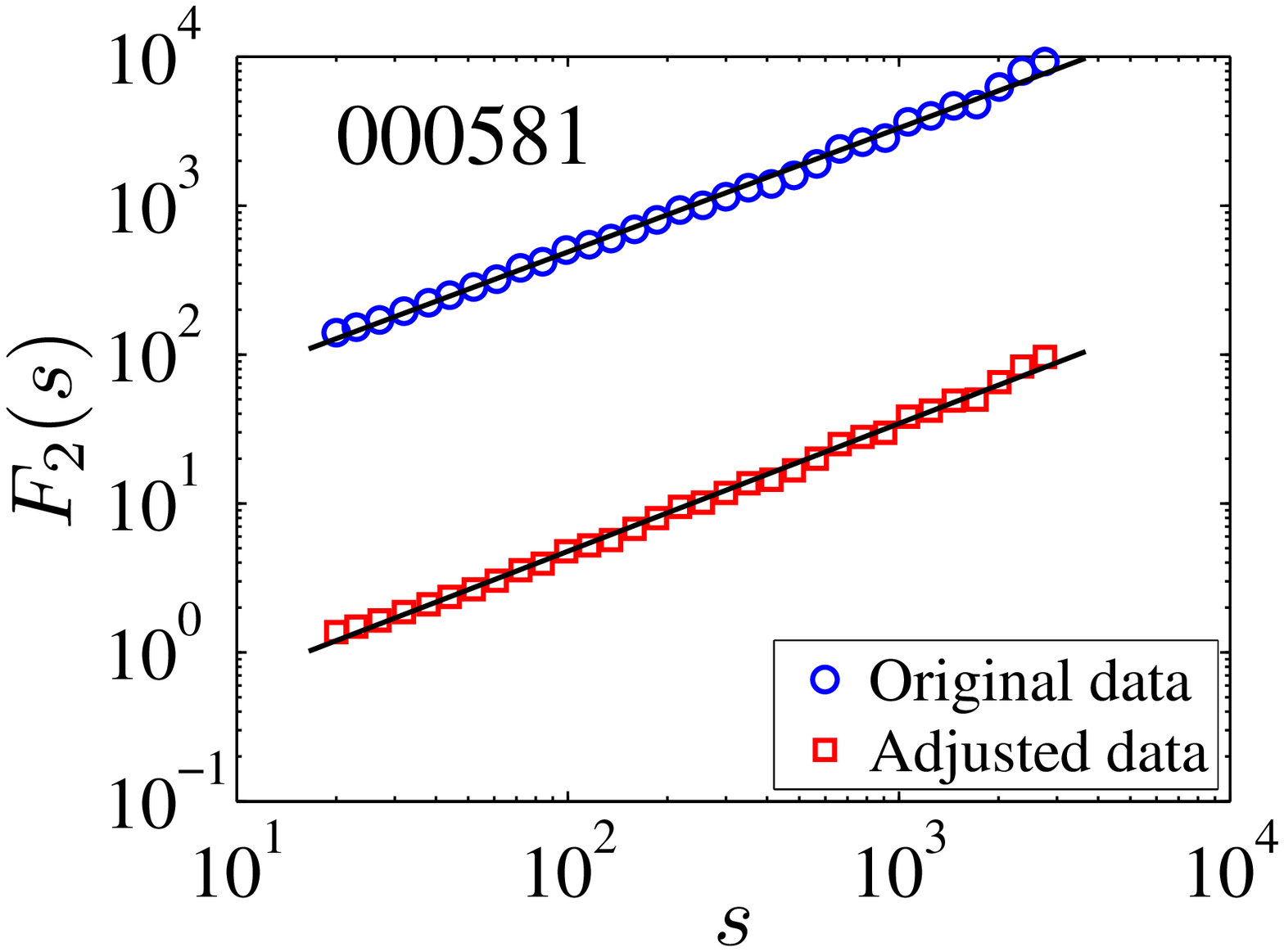}
\includegraphics[width=4cm]{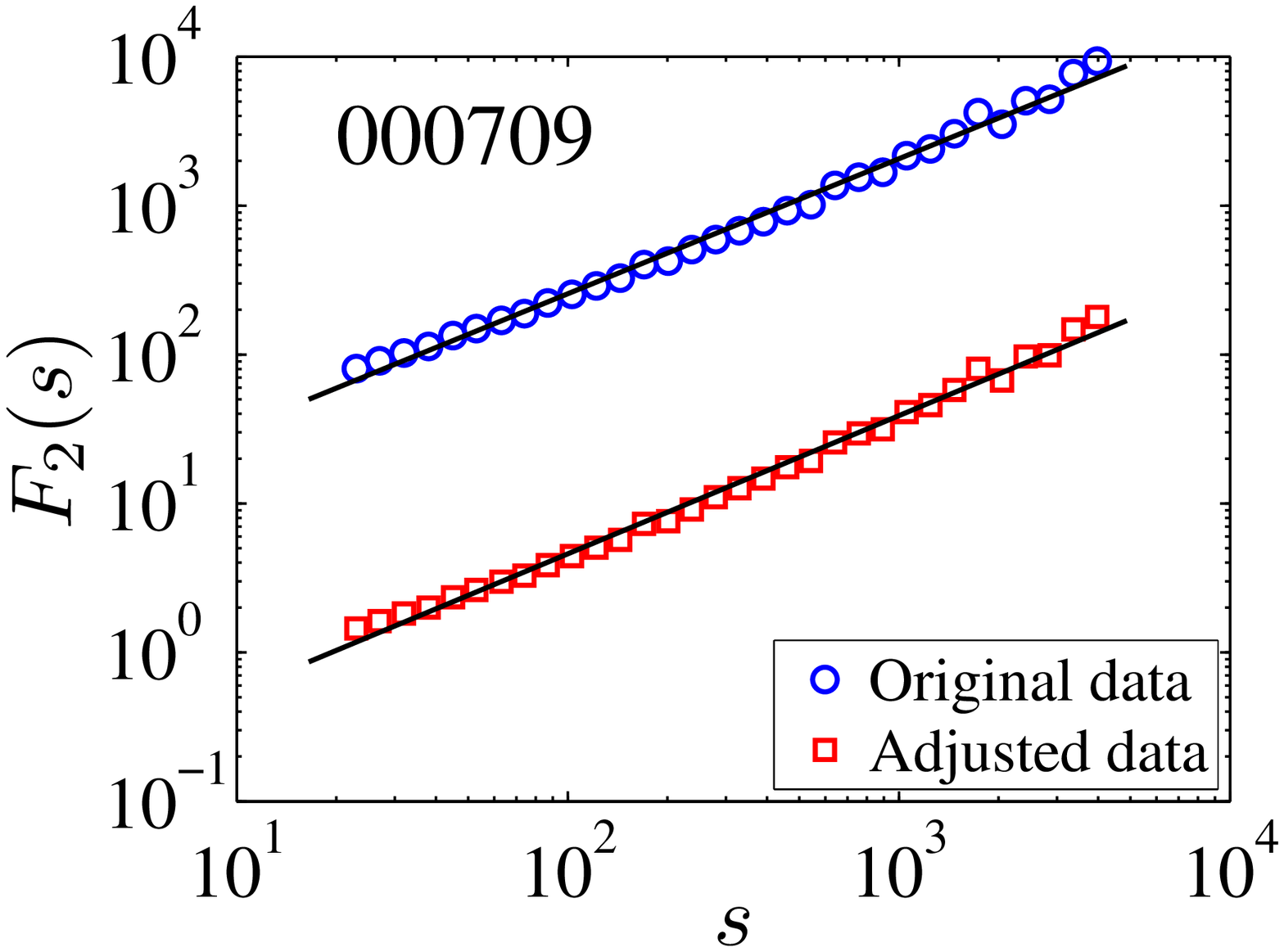}
\caption{\label{Fig:F2q}(Color online) Log-log plots of the overall fluctuation function $F_2(s)$ with respect to the interval size $s$ for four different stocks. The open circles and squares stand for the original and adjusted data, respectively. The solid lines are the power-law fits to the data. The curves for the adjusted data have been shifted downwards for clarity.}
\end{figure}

Figure \ref{Fig:F2q} shows the log-log plots of the overall fluctuations $F_2(s)$ as a function of the interval size $s$ for the four randomly selected stocks. Excellent power-law dependence is observed for each curve with the scaling range spanning more than two orders of magnitude. No clear crossover is observed in the present case, which is however quite common for other financial quantities \cite{Hu-Ivanov-Chen-Carpena-Stanley-2001-PRE,Chen-Ivanov-Hu-Stanley-2002-PRE,Chen-Hu-Carpena-Bernaola-Galvan-Stanley-Ivanov-2005-PRE}. We also find that the two curves in each plot parallel to each other, which means that the intraday patterns have minor influence on the long-term power-law correlations.

In Table \ref{Tb:DFA_F2q}, the DFA scaling exponents $H$ for all the 22 stocks are listed. All the DFA scaling exponents are significantly greater than 0.5, indicating that the inter-cancelation durations have strong long memory. The DFA scaling exponents $H$ of both the original data and the adjusted data do not differ much, which confirms that the intraday patterns have little influence on the long-range dependence of inter-cancelation durations. This finding provides further evidence that the cancelation process is a non-Poisson process.

\begin{table}[htp]
\centering
\caption{Characteristic parameters of the long-term memory and the multifractal nature of the inter-cancelation durations. The first column gives the stock codes. The second column lists the mean number of cancelations $\langle N_T\rangle$ in one trading day. The third and the fifth columns list the DFA scaling exponents $H_1$ of the original data and $H_2$ of the adjusted data.  The rest two columns list the width of multifractal spectrum $\Delta\alpha$ of the inter-cancelation durations for the 22 stocks. }
\medskip
\label{Tb:DFA_F2q} \centering
\begin{tabular}{rrrcrc}
 \hline \hline
  \multicolumn{1}{c}{Code} &
  \multicolumn{1}{c}{$\langle N_T\rangle$} &
  \multicolumn{2}{c}{\begin{tabular}{cc} \multicolumn{2}{c}{Original data} \\ \hline ~~~~~$H_1$~~~~~ & ~~~~~$\Delta\alpha$
  \end{tabular}} &
  \multicolumn{2}{c}{\begin{tabular}{cc} \multicolumn{2}{c}{Adjusted data} \\ \hline ~~~~~$H_2$~~~~~ & ~~~~~$\Delta\alpha$
  \end{tabular}}\\
  \hline
    000001 &    2476 &    0.91 $\pm$    0.02 &    0.76 &    0.94 $\pm$    0.00 &    0.88\\
    000002 &     327 &    0.93 $\pm$    0.02 &    0.76 &    0.96 $\pm$    0.00 &    0.89\\
    000009 &    1546 &    0.97 $\pm$    0.05 &    0.53 &    1.01 $\pm$    0.00 &    0.49\\
    000012 &     928 &    0.83 $\pm$    0.03 &    0.58 &    0.86 $\pm$    0.00 &    0.59\\
    000016 &     497 &    0.89 $\pm$    0.04 &    0.67 &    0.92 $\pm$    0.00 &    0.60\\
    000021 &    1291 &    0.90 $\pm$    0.02 &    0.74 &    0.93 $\pm$    0.00 &    0.69\\
    000024 &     370 &    0.90 $\pm$    0.03 &    0.54 &    0.93 $\pm$    0.00 &    0.46\\
    000027 &     300 &    0.90 $\pm$    0.02 &    0.72 &    0.94 $\pm$    0.00 &    0.71\\
    000063 &     254 &    0.89 $\pm$    0.05 &    0.74 &    0.91 $\pm$    0.00 &    0.70\\
    000066 &     903 &    0.89 $\pm$    0.04 &    0.59 &    0.92 $\pm$    0.00 &    0.53\\
    000088 &      57 &    0.80 $\pm$    0.17 &    0.61 &    0.83 $\pm$    0.00 &    0.54\\
    000089 &     185 &    0.92 $\pm$    0.04 &    0.47 &    0.95 $\pm$    0.00 &    0.39\\
    000429 &     305 &    0.85 $\pm$    0.03 &    0.82 &    0.88 $\pm$    0.00 &    0.69\\
    000488 &     277 &    0.88 $\pm$    0.07 &    0.59 &    0.91 $\pm$    0.00 &    0.43\\
    000539 &     226 &    0.86 $\pm$    0.09 &    0.81 &    0.88 $\pm$    0.00 &    0.76\\
    000541 &     164 &    0.85 $\pm$    0.06 &    0.60 &    0.89 $\pm$    0.00 &    0.57\\
    000550 &    1051 &    0.87 $\pm$    0.05 &    0.73 &    0.90 $\pm$    0.00 &    0.68\\
    000581 &     236 &    0.84 $\pm$    0.05 &    0.89 &    0.87 $\pm$    0.00 &    0.76\\
    000625 &    1071 &    0.89 $\pm$    0.05 &    0.75 &    0.91 $\pm$    0.00 &    0.65\\
    000709 &     542 &    0.92 $\pm$    0.05 &    0.69 &    0.94 $\pm$    0.00 &    0.71\\
    000720 &     129 &    0.84 $\pm$    0.20 &    0.74 &    0.85 $\pm$    0.00 &    0.69\\
    000778 &     379 &    0.92 $\pm$    0.05 &    0.64 &    0.95 $\pm$    0.00 &    0.55\\
  \hline \hline
\end{tabular}
\end{table}

\subsection{Multifractal nature}
\label{S2:MFDFA}

We now investigate the multifractal nature of inter-cancelation durations by using the multifractal detrended fluctuation analysis (MF-DFA)
\cite{Kantelhardt-Zschiegner-KoscielnyBunde-Havlin-Bunde-Stanley-2002-PA}. The first several steps of the MF-DFA procedure are essentially
the same as the DFA procedure, and the multifractal detrended fluctuation is defined as the following form,
\begin{equation}
 F_q(s) =\left\{\frac{1}{2N_s}\sum_{k=1}^{2N_s}[r_k(s)]^q \right\}^\frac{1}{q}, \label{Eq:MFDFA}
\end{equation}
where the moment order $q$ varies in real number except for $q = 0$ in which $F_0(s)$ is defined as
\begin{equation}
 F_0(s) =\exp\left\{\frac{1}{2N_s}\sum_{k=1}^{2N_s}ln[f_k(s)] \right\}. \label{Eq:MFDFA:q0}
\end{equation}
For each given value $q$, we vary the value of $s$ in the interval $[20, N/4]$ as in DFA. If the series is long-range power correlated, a power-law relation is expected between the detrended fluctuation function $F_q(s)$ and the size $s$,
\begin{equation}
 F_q(s) \sim s^{h(q)}, \label{Eq:h:q0}
\end{equation}
where $h(q)$ is the MF-DFA scaling exponent. Note that when $q = 2$, $h(2)$ is nothing but the DFA scaling exponent $H$. The singularity strength $\alpha$ and its spectrum $f(\alpha)$ can be obtained based on the Legendre transform,
\begin{equation}
 \left\{
 \begin{array}{rcl}
 \alpha&=&qh'(q)+h(q)~~\\
 f(\alpha)&=&q^2h'(q)+1
 \end{array}\right..
 \label{Eq:Null}
\end{equation}

To investigate the multifractality of the inter-cancelation durations, the fluctuation functions $F_q(s)$ are calculated for each stock. Since the size of each time series is finite, the estimate of $F_q(s)$ will fluctuate remarkably for large values of $|q|$, especially when the scale $s$ is large. We focus on $q \in [-6, 6]$ to obtain reasonable statistics in the estimation of $F_q(s)$. We show the detrended fluctuation functions of the original data and the adjusted data for the four typical stocks. Again, nice power laws are observed for all the curves.

\begin{figure}[htb]
\centering
\includegraphics[width=4cm]{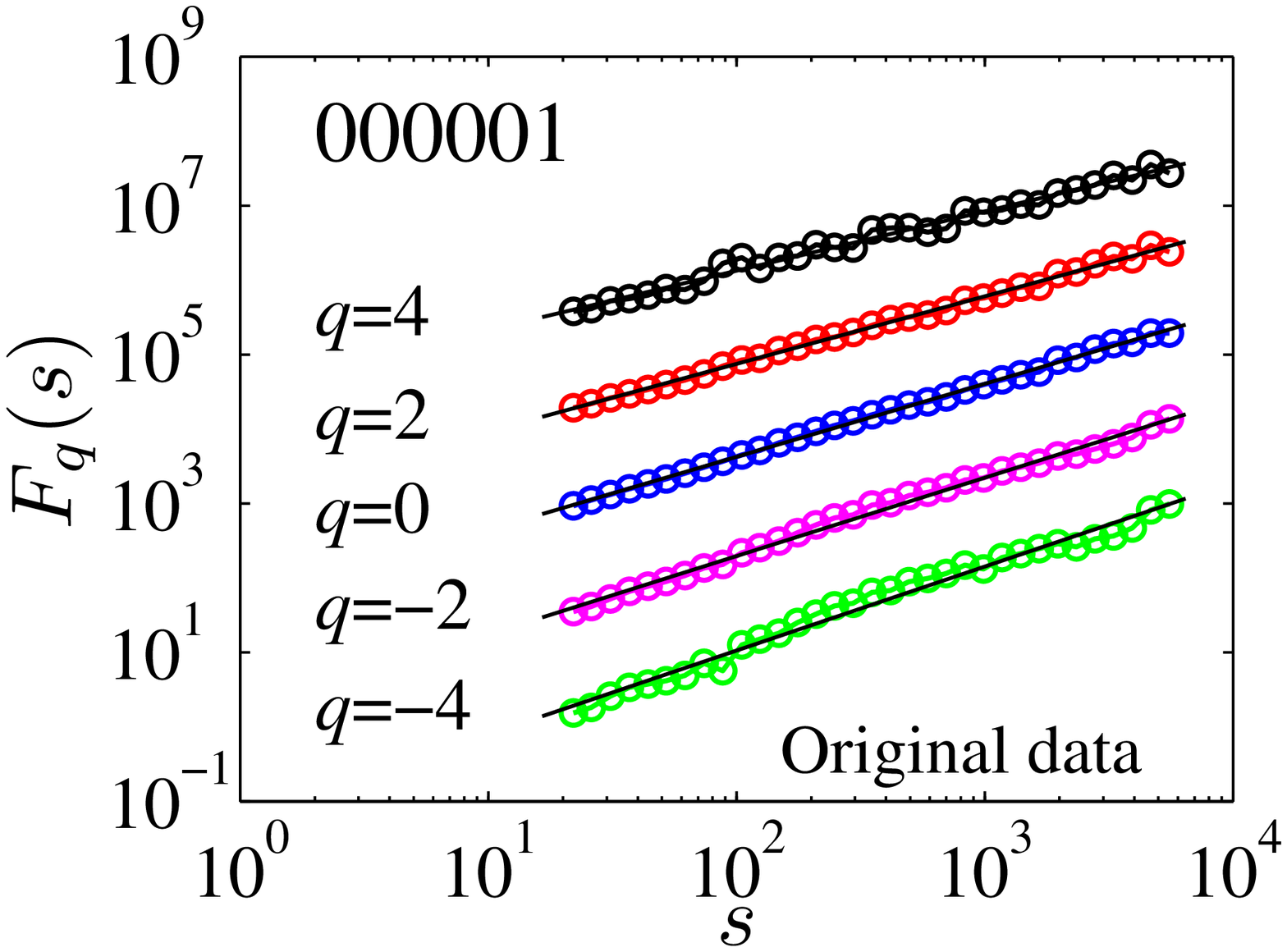}
\includegraphics[width=4cm]{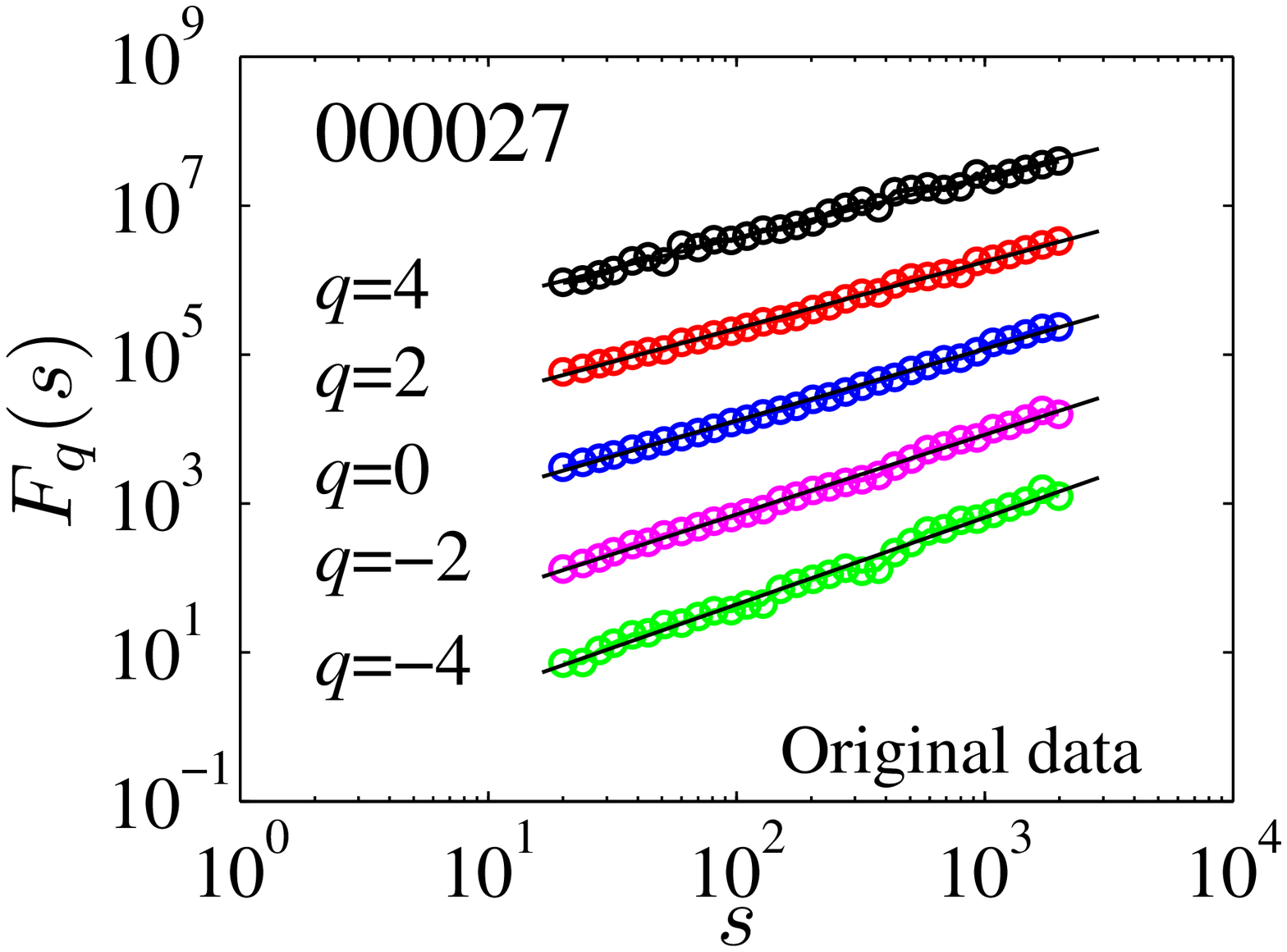}
\includegraphics[width=4cm]{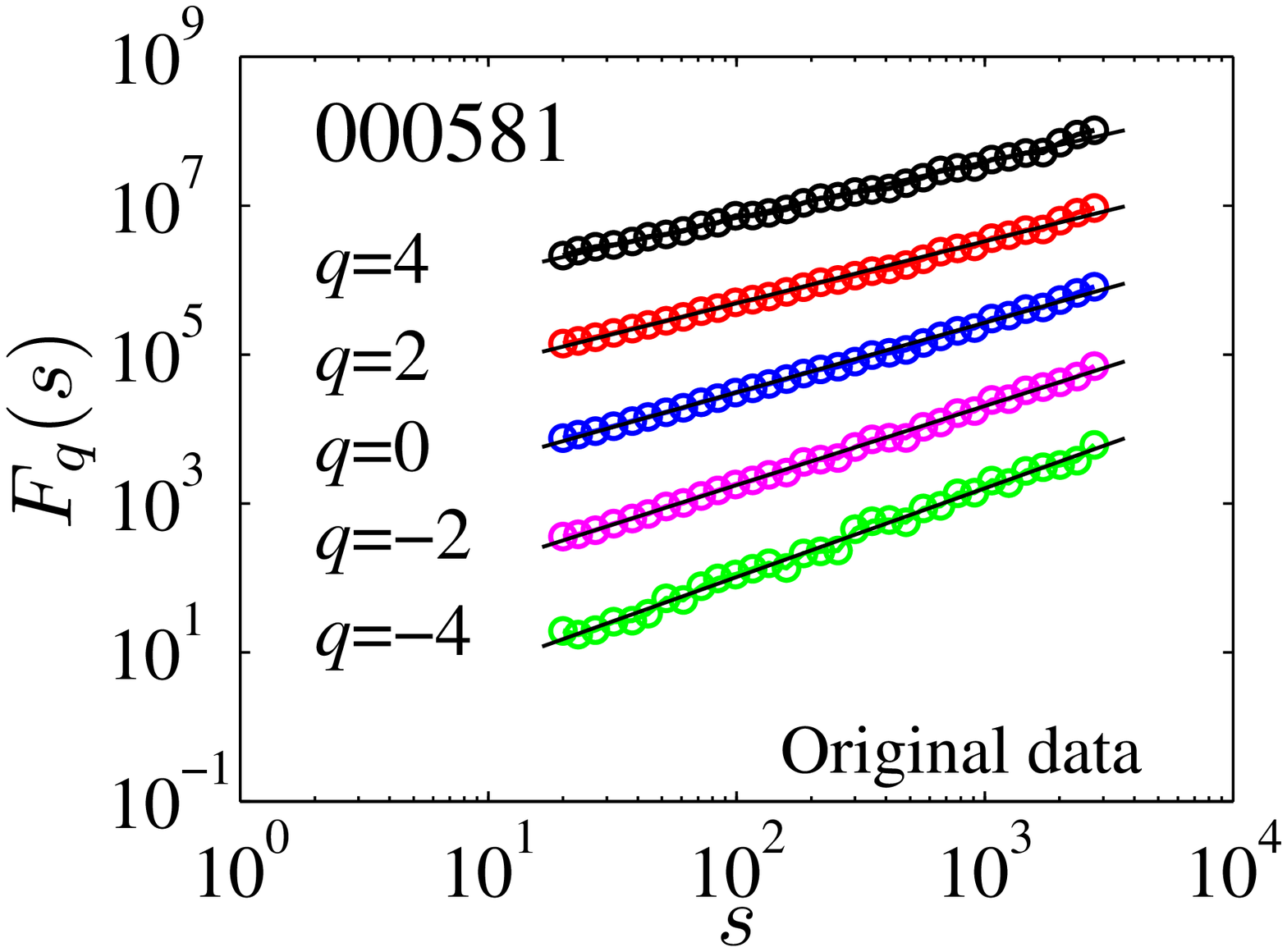}
\includegraphics[width=4cm]{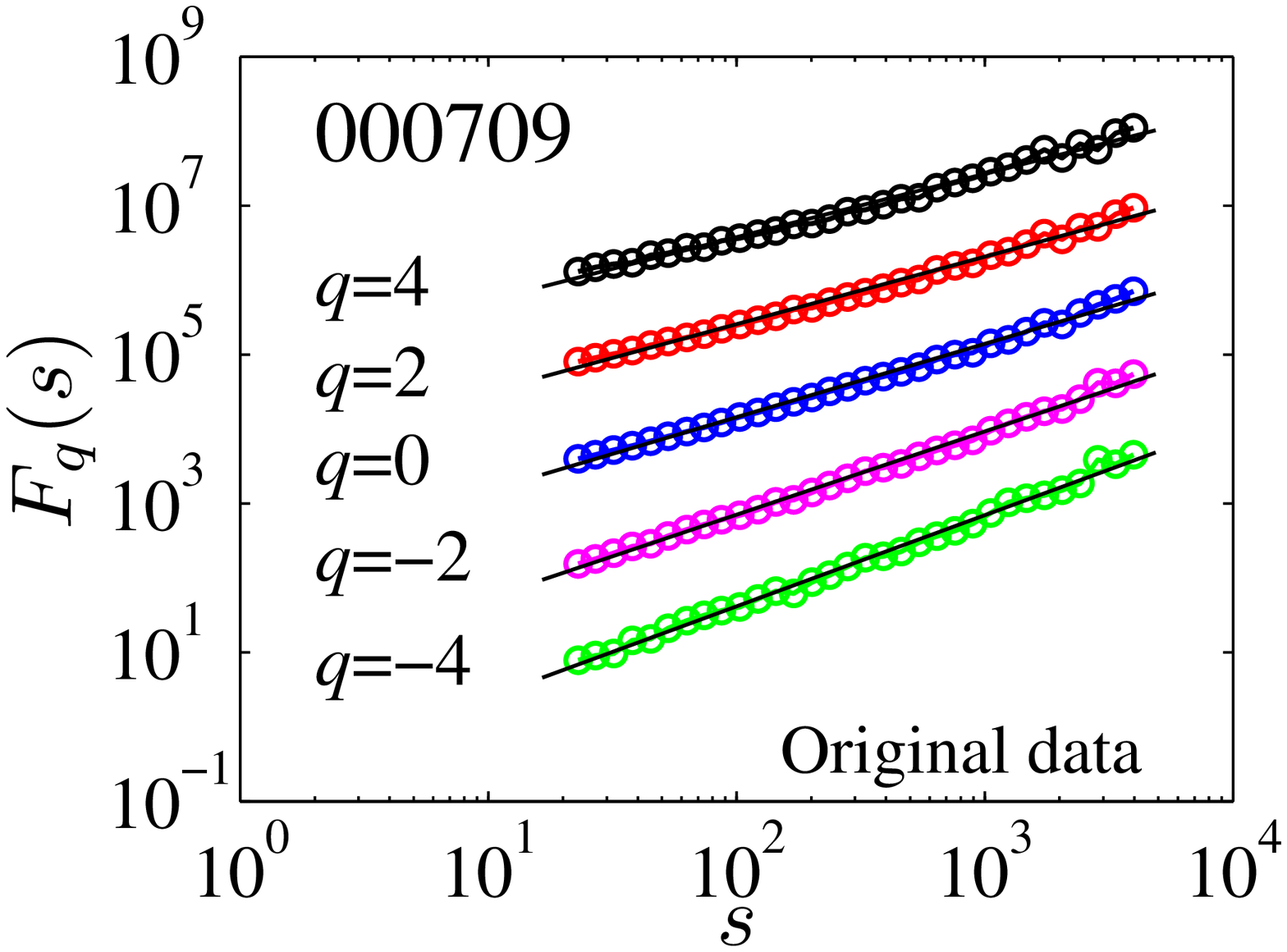}
\includegraphics[width=4cm]{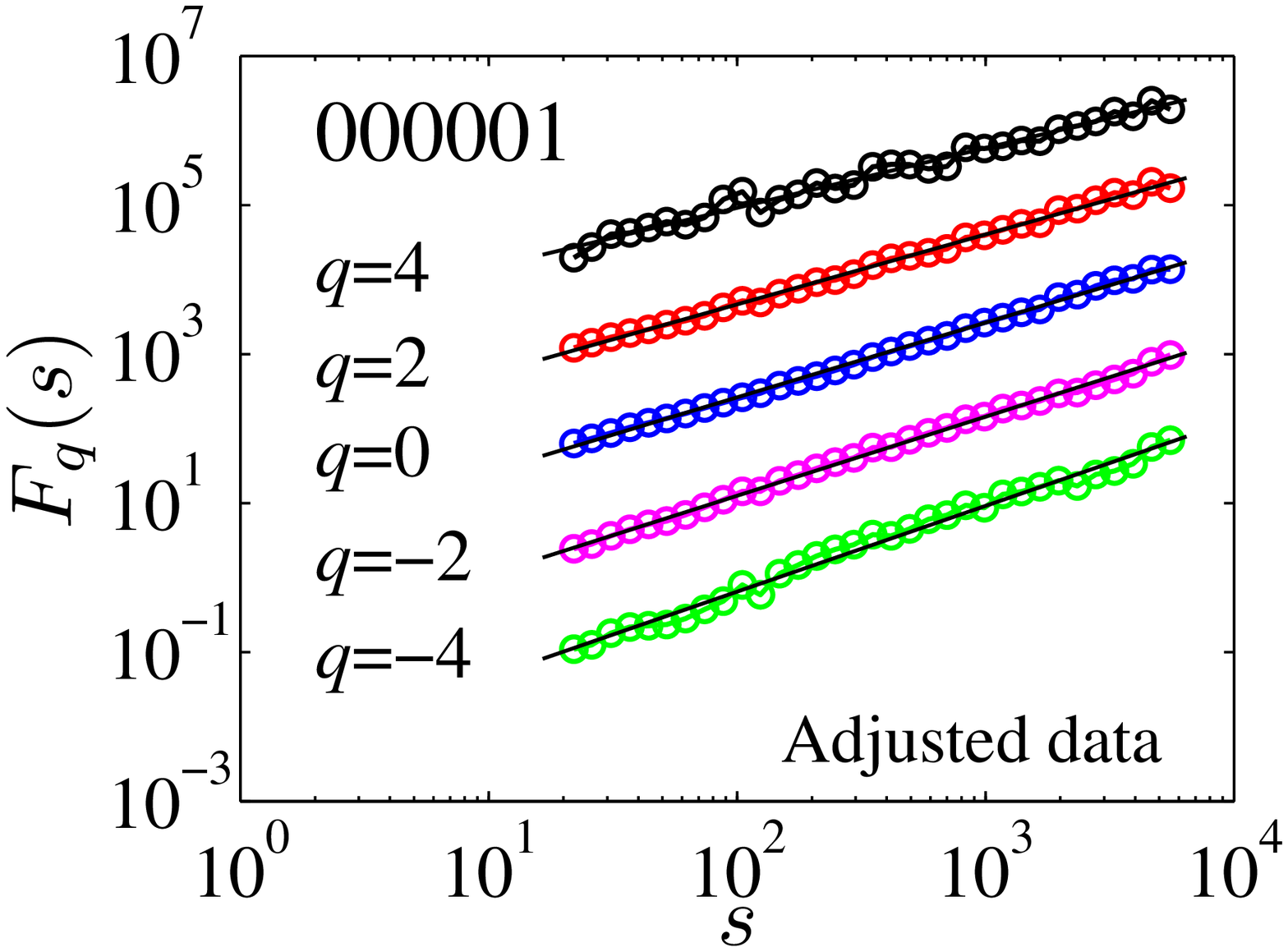}
\includegraphics[width=4cm]{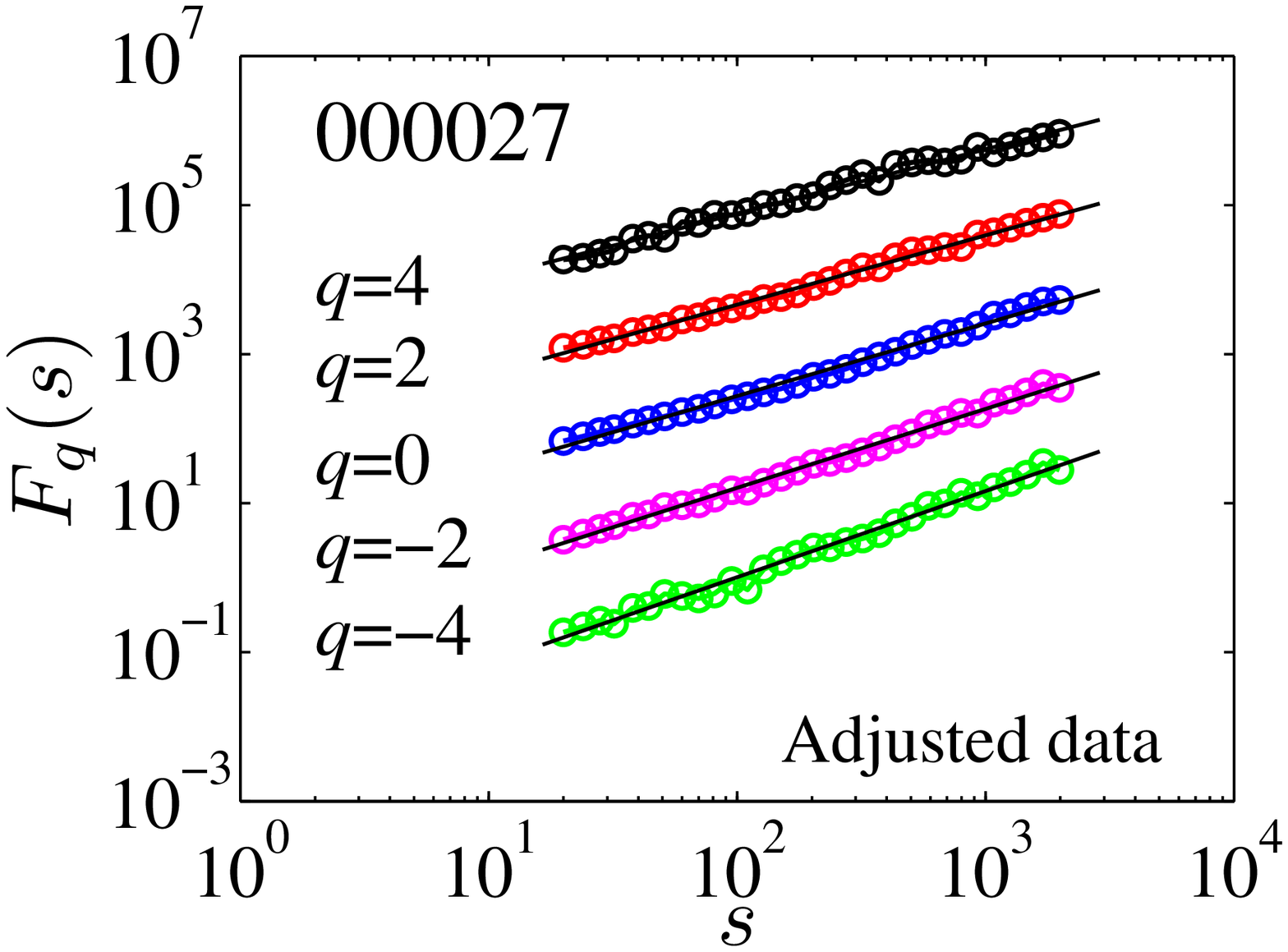}
\includegraphics[width=4cm]{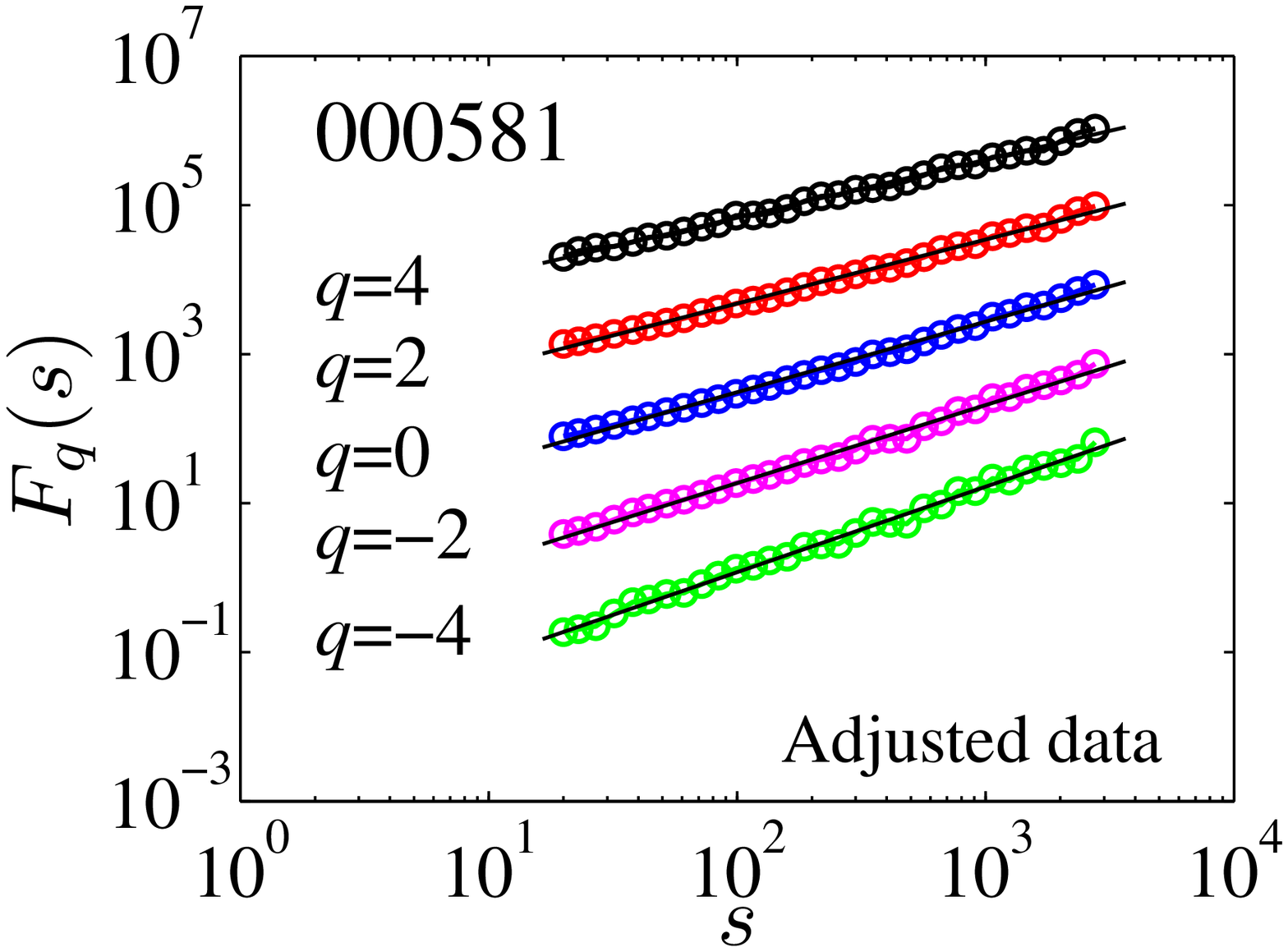}
\includegraphics[width=4cm]{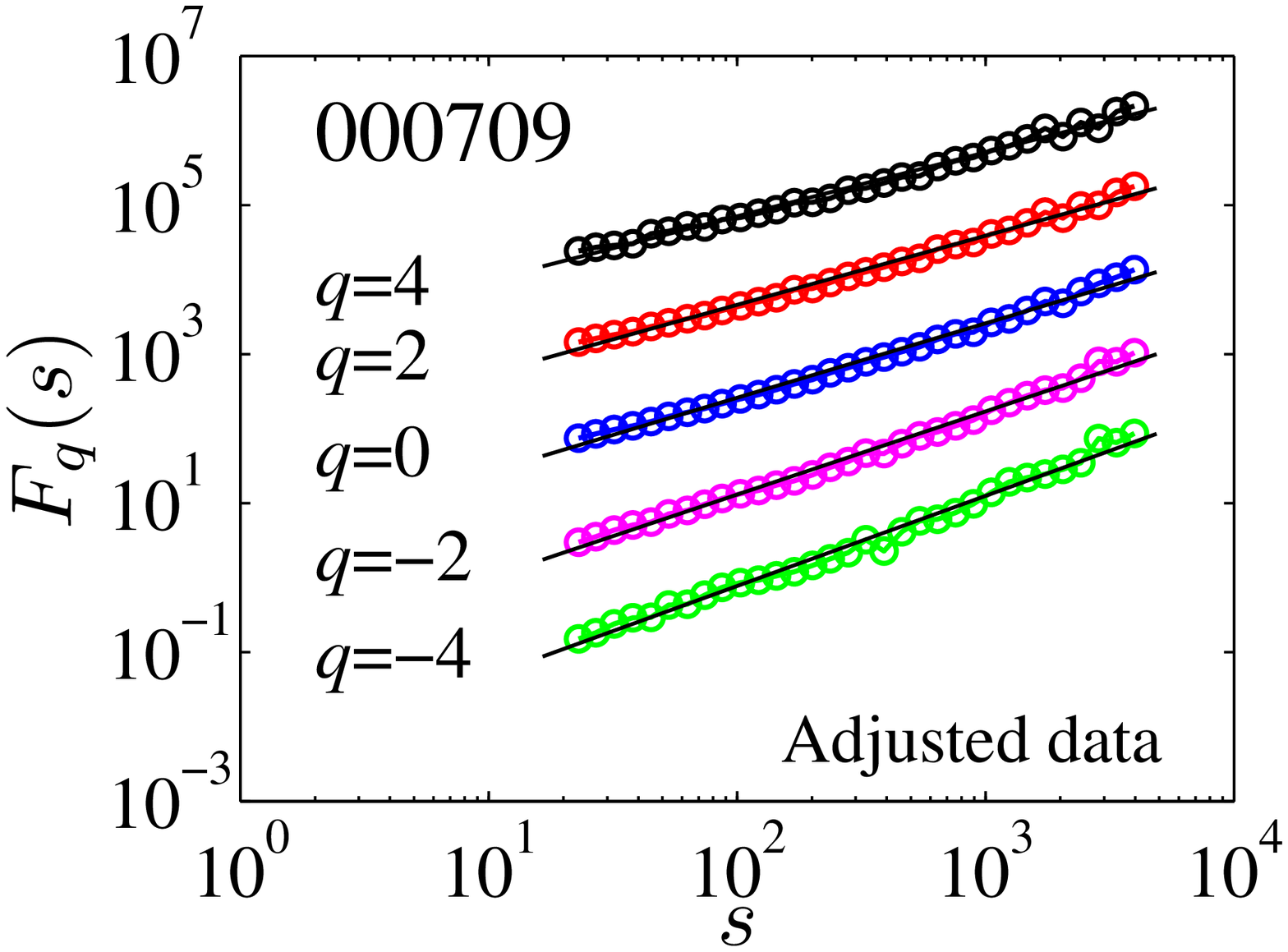}
\caption{\label{Fig:F5q} (Color online) Dependence of the detrended fluctuation functions $F_q(s)$ of the original data (upper panel) and the adjusted data (lower panel) with respect to the scale $s$ for the four randomly chosen stocks. The solid lines are the best power-law fits to the data in the corresponding scaling ranges.}
\end{figure}

\begin{figure}[htb]
\centering
\includegraphics[width=7cm]{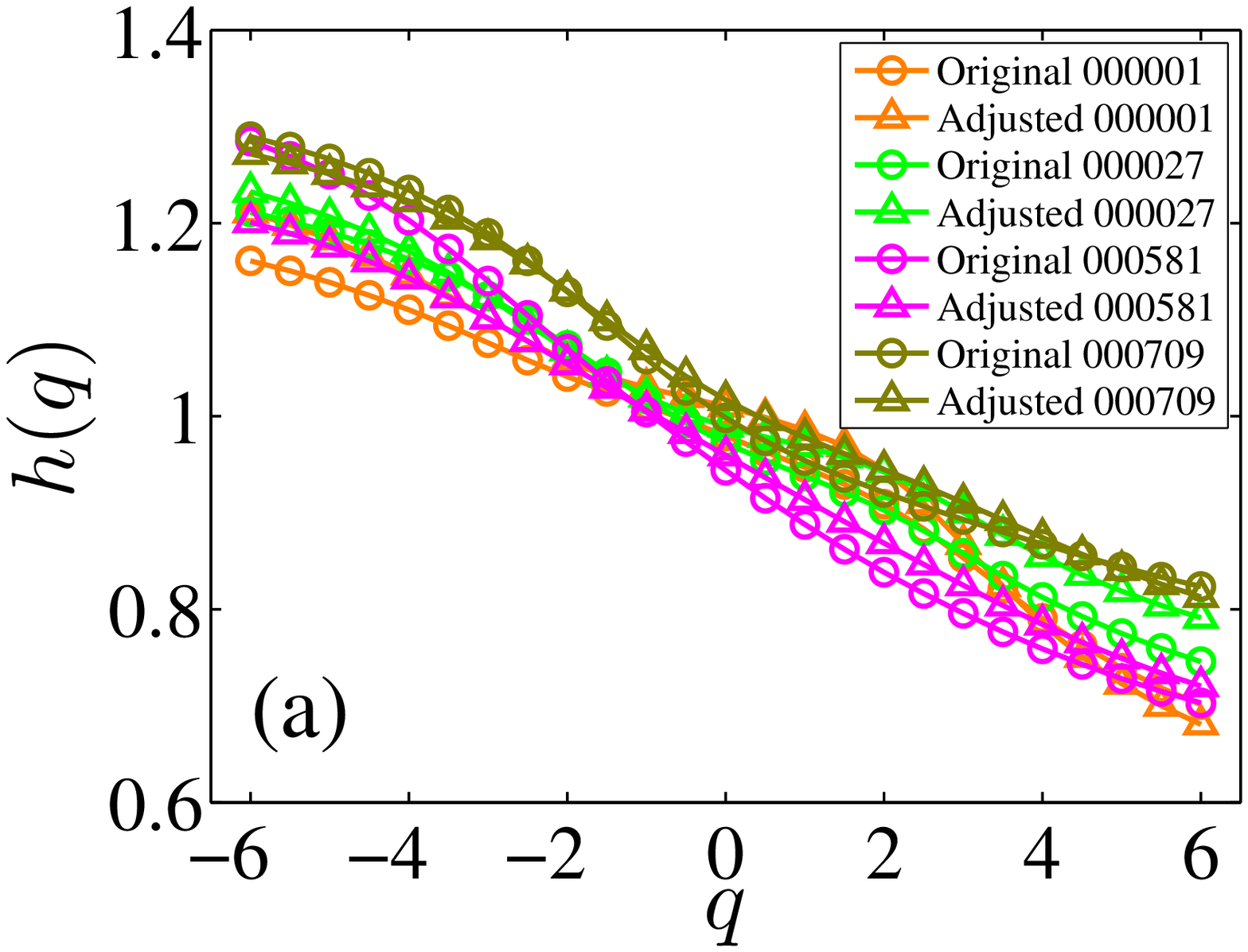}
\includegraphics[width=7cm]{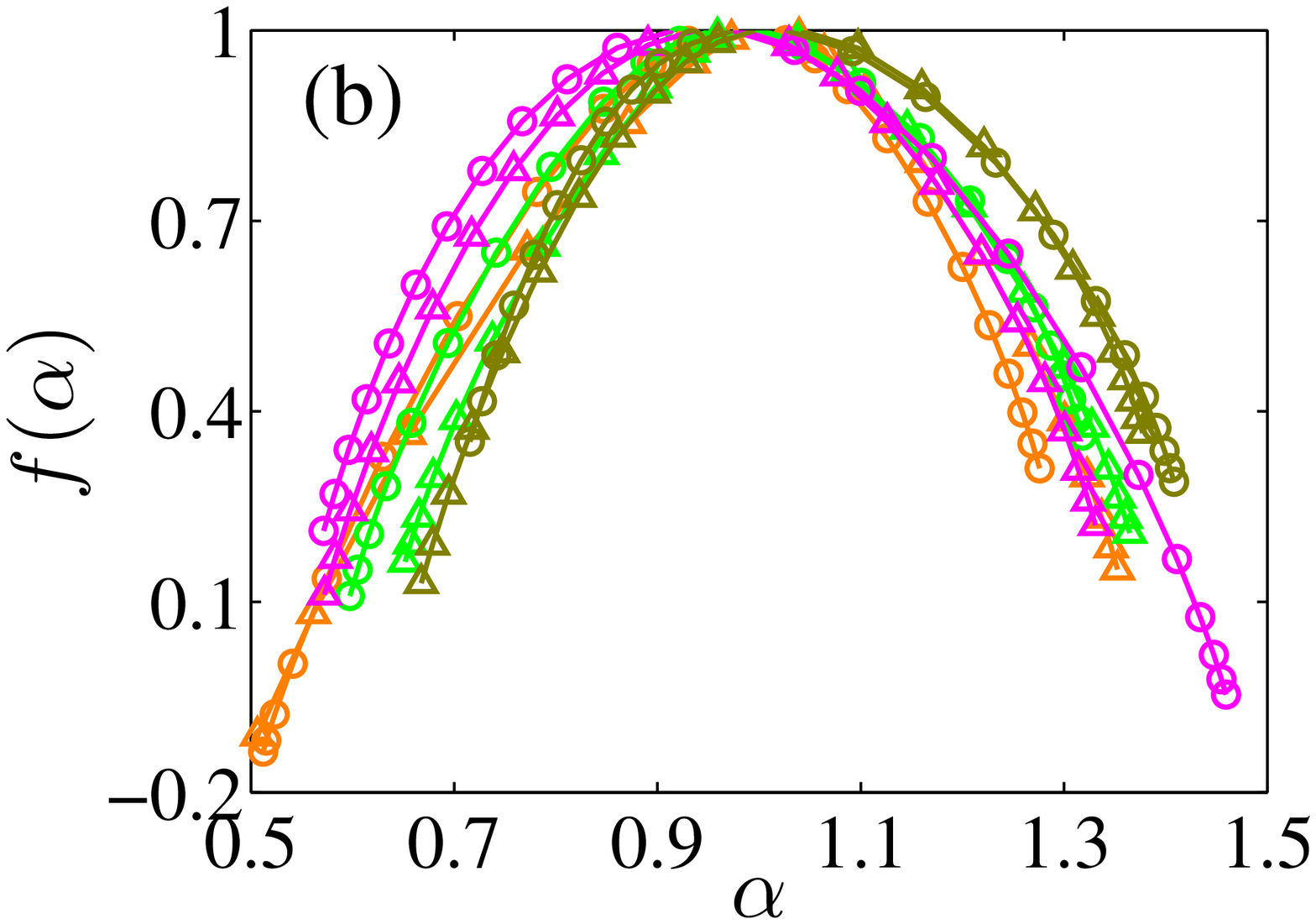}
\caption{\label{Fig:MFDFA}(Color online) Multifractal analysis of the inter-cancelation durations for four typical stocks. Panel (a) shows the
MF-DFA scaling exponent $h(q)$ with respect to $q$, and panel (b) shows the corresponding multifractal spectra $f(\alpha)$. }
\end{figure}

For each given $q$, the MF-DFA scaling exponent $h(q)$ can be estimated by the linear regression between $\ln[F_q(s)]$ and $\ln s$. The resultant $h(q)$ functions are shown in Fig.~\ref{Fig:MFDFA}(a), which decrease with the increase of $q$. Figure \ref{Fig:MFDFA}(b) illustrates the multifractal spectra for the four stocks. It is evident that the inter-cancelation durations possess multifractal nature, on which the intraday patterns have minor impacts.

The strength of the multifractal nature can be quantified by the width of the singularity spectrum
\begin{equation}
  \Delta\alpha \triangleq \alpha_{\rm{max}} -\alpha_{\rm{min}}.
\end{equation}
A large value of $\Delta\alpha$ corresponds to stronger multifractality. The values of the singularity width $\Delta\alpha$ for the 22 stocks are calculated and listed in Table \ref{Tb:DFA_F2q}. These $\Delta\alpha$ values are significantly different from zero as a hallmark of strong multifractality.

\section{Conclusion}
\label{S5:conclusion}

We have investigated the inter-cancelation durations calculated from the limit order book data of 22 liquid Chinese stocks traded on the SZSE in the whole year 2003. It is found that the probability densities decrease with the increasing of the inter-cancelation durations. The 22 empirical densities of the rescaled durations collapse onto a single curve, showing a nice scaling pattern. This scaling behavior is also observed in the distributions of buyer-initiated inter-cancelation durations and seller-initiated inter-cancelation durations, which implies that there are common features in the cancelation behavior of market participants.

We then model the three ensemble densities of rescaled inter-cancelation durations for all cancelations, buyer-initialed cancelations and seller-initialed cancelations by using the Weibull and the Tsallis $q$-exponential. By using the maximum likelihood estimation and nonlinear least-squares estimation methods, it is found that all the three kinds of inter-cancelation durations can be well modeled by the Weibull function. We also study the conditional distribution of rescaled inter-cancelation durations which follows a certain set of rescaled inter-cancelation durations. It is found that large (resp. small) durations tend to follow large (resp.) durations. It means that the inter-cancelations exhibit short-term memory.

In addition, the intraday pattern, the long memory, and the multifractal nature of the durations are also investigated. An inverse $U$-shaped intraday pattern is revealed for each stock. The DFA and the MF-DFA methods are applied to both the original data and the adjusted data after removing the intraday patterns. We show that the intraday patterns have minor influence on these correlation properties. The results show that the durations possess both the long memory and the multifractal nature.

\bigskip
{\textbf{Acknowledgments:}}

This work was partially supported by the ``Shu Guang'' project (Grant No. 2008SG29) and the ``Chen Guang'' project (Grant No. 2008CG37) sponsored by Shanghai Municipal Education Commission and Shanghai Education Development Foundation, the Zhejiang Provincial Natural Science Foundation of
China (Grant No. Z6090130), and the Program for New Century Excellent Talents in University (Grant No. NCET-07-0288).

%\pagebreak
%\bibliography{Bibliography}
\bibliography{E:/Papers/Auxiliary/Bibliography}

\end{document}